\documentclass[authoryear]{article}

\usepackage{geometry}
\usepackage[utf8]{inputenc}
\usepackage[T1]{fontenc}
\usepackage[onehalfspacing]{setspace}
\usepackage{charter}
\usepackage[charter]{mathdesign}

\usepackage{graphicx}
\usepackage{bm}
\usepackage[usenames]{color}
\usepackage{longtable}
\usepackage{multirow}
\usepackage{multicol}
\usepackage[dvipsnames,svgnames,table]{xcolor}
\usepackage{lscape}
\usepackage{pdflscape}
\usepackage{authblk}
\usepackage{textcomp}
\usepackage[T1]{fontenc}
\usepackage{setspace}
\usepackage{hyperref}
\hypersetup{colorlinks=true,linkcolor=blue,citecolor=blue}
\usepackage{lineno}
\usepackage[ruled]{algorithm2e}
\usepackage{url}
\usepackage{tabularx,longtable,supertabular,booktabs,dcolumn,multicol,multirow}
\usepackage[hang, singlelinecheck=off, center]{caption}
\usepackage[flushleft]{threeparttable}
\usepackage{adjustbox}
\usepackage{subcaption}
\usepackage{amsthm}
\usepackage{amsmath}
\usepackage{tabulary}
\usepackage{ragged2e}

\usepackage{siunitx}
\usepackage{booktabs}
\usepackage[title]{appendix}

\usepackage[round, authoryear]{natbib}
\usepackage{changepage}
\usepackage{array}

\begin{document}

\begin{spacing}{1.2}
\begin{flushleft}
\huge \textbf{A Generalized Continuous-Multinomial Response Model with a t-distributed Error Kernel} \\
\vspace{\baselineskip}
\normalsize
January 18, 2020 \\
\vspace{\baselineskip}
Subodh Dubey\textsuperscript{*} \\
Department of Transport and Planning \\
Delft University of Technology, Netherlands\\
S.K.Dubey@tudelft.nl\\
\vspace{\baselineskip}
Prateek Bansal\textsuperscript{*} \\
\textbf{(Corresponding author)}\\
School of Civil and Environmental Engineering \\
Cornell University, United States \\
pb422@cornell.edu \\
\vspace{\baselineskip}
Ricardo A. Daziano \\
School of Civil and Environmental Engineering \\
Cornell University, United States  \\
daziano@cornell.edu \\
\vspace{\baselineskip}
Erick Guerra\\
Department of City and Regional Planning \\
University of Pennsylvania, United States\\
erickg@upenn.edu \\
\vspace{\baselineskip}
\textsuperscript{*} These authors contributed equally to this work.
\end{flushleft}
\end{spacing}

\newpage
\section*{Abstract}
In multinomial response models, idiosyncratic variations in the indirect utility are generally modeled using Gumbel or normal distributions. This study makes a strong case to substitute these thin-tailed distributions with a t-distribution. First, we demonstrate that a model with a t-distributed error kernel better estimates and predicts preferences, especially in class-imbalanced datasets. Our proposed specification also implicitly accounts for decision-uncertainty behavior, i.e. the degree of certainty that decision-makers hold in their choices relative to the variation in the indirect utility of any alternative. Second -- after applying a t-distributed error kernel in a multinomial response model for the first time -- we extend this specification to a generalized continuous-multinomial (GCM) model and derive its full-information maximum likelihood estimation procedure. The likelihood involves an open-form expression of the cumulative density function of the multivariate t-distribution, which we propose to compute using a combination of the composite marginal likelihood method and the separation-of-variables approach. Third, we establish finite sample properties of the GCM model with a t-distributed error kernel (GCM-t) and highlight its superiority over the GCM model with a normally-distributed error kernel (GCM-N) in a Monte Carlo study. Finally, we compare GCM-t and GCM-N in an empirical setting related to preferences for electric vehicles (EVs). We observe that accounting for decision-uncertainty behavior in GCM-t results in lower elasticity estimates and a higher willingness to pay for improving the EV attributes than those of the GCM-N model. These differences are relevant in making policies to expedite the adoption of EVs.

\newpage

\section{Introduction}
Discrete-response (e.g., count, ordered, binary, multinomial) models are popular across various disciplines such as applied economics, transportation, marketing, and political science. In these models, error structures are specified at different modeling levels because the researcher does not have full information about the data generating process (DGP). For instance, the total indirect utility in additive random utility maximization (ARUM) \citep{mcfadden1973conditional} based choice models is specified as the sum of a deterministic index (depending on observables) and a random error term or taste shock. Such error specifications in empirical studies are generally governed by ease of estimation rather than structural appropriateness \citep{vijverberg2016pregibit}. In particular, without worrying about the characteristics of the data in hand, Gumbel/extreme-value (logit) or normal (probit) error kernels are commonly used. These traditional logit and probit links have been replaced by a t-distributed error kernel, i.e. a robit link \citep{liu2004robit}, in multi-level modelling applications to handle heavy-tailed error distributions. Moreover, a t-distribution with an estimable degree-of-freedom (DOF) actually generalizes logit and probit.\footnote{The t-distribution with about seven and \textit{a large} (above thirty) DOF approximates logistic and normal distributions, respectively.} However, we are not aware of the use of the robit link in modeling multinomial responses, perhaps because the benefits of the generalization are unclear and the estimation of the resulting model is cumbersome.  

In this study, we present the  first application of a multinomial response model with a t-distributed error kernel, multinomial robit (MNR) model henceforth. Whereas the proposed MNR model retains all merits of the multinomial probit (MNP) model\footnote{Similar to MNP, MNR allows for flexible substitution patterns -- correlations across indirect utilities of alternatives -- without necessarily including variation in parameters across decision-makers.}, we illustrate three additional advantages of adopting MNR in practice. First, we highlight that the robit link is superior to the probit link in estimating and predicting preferences in unbalanced datasets where one or more alternatives have small shares. Second, we parameterize the DOF of the t-distributed error kernel as a function of demographics and show how this specification can help in capturing decision uncertainty of decision-makers that standard compensatory ARUM models cannot account for. Third, if the error distribution in the true DGP is robit (symmetric heavy-tailed), probit fails to recover the true model parameters. We numerically establish this benefit of MNR over MNP in a Monte Carlo study. Given the growing interest in the joint modeling of mixed datasets across multiple disciplines \citep[see][for applications]{de2013analysis}, we further extend MNR to a generalized continuous-multinomial (GCM) response model with a t-distributed error kernel that facilitates simultaneous consideration of multiple multinomial and multiple continuous dependent variables. We note that the advantages of considering a t-distributed (over a normally-distributed) error kernel are even more evident in the GCM response model because heavy-tailed distributions are more commonly observed in continuous outcomes. Moreover, the joint GCM model offers several advantages, namely: a) statistically efficient estimation, b) easier hypothesis testing and better power of statistical tests, c) avoidance of inconsistencies in a situation when continuous and multinomial outcomes have structural dependence or have common unobserved factors \citep[see][for a detailed discussion]{bhat2015introducing}. For example, the GCM response model is appropriate for the joint modeling of commute distance (continuous variable) and choice of residential location (multinominal variable) because both outcomes may affect each other and may share common unobserved factors. Joint modeling is thus desirable, but it is generally intractable due to an absence of convenient distributions to represent the conditional and/or joint relationship between the outcomes. The proposed GCM response model with a t-distributed error kernel (henceforth, GCM-t) could not have been estimated conveniently without an elegant statistical property of the t-distribution: the conditional distribution of a joint/multivariate t-distribution is also a t-distribution \citep{ding2016conditional}. 

The contribution of this study is thus threefold. First, we illustrate the importance of t-distributed error kernels in multinomial choice modeling. Second, we derive a full-information maximum likelihood procedure to estimate MNR and GCM-t models. The likelihood expressions of both models involve evaluation of high-dimensional multi-variate-t-cumulative density (MVTCD) functions. We adopt the composite marginal likelihood (or paired-likelihood) approach to first decompose the multi-dimensional MVTNCD integral into multiple pairs and thus reduce the integral dimensionality \citep[see][]{varin2011overview,xu2011robustness}. Subsequently, similar to the Geweke-Hajivassiliou-Keane (GHK) method to simulate multivariate normal cumulative density functions \citep{genz1992numerical,hajivassiliou1996simulation}, we use a separation-of-variables approach to simulate MVTCD functions \citep{genz1999numerical}. Third, we numerically verify the statistical properties of the maximum likelihood estimator of the GCM-t model in a Monte Carlo study. We also compare the performance of GCM-t and GCM with normally-distributed error kernel (GCM-N) models in a second simulation study. Finally, we validate the simulation results and highlight the advantages of the robit link (over probit) in an empirical study in the context of policies for on-street parking with charging facilities for electric vehicles. The empirical data looks into the adoption of electric vehicles and the outcomes of interest are a household's vehicle miles traveled (continuous) and vehicle-purchase preferences (multinomial).

The remainder of this paper is organized as follows. We present the contextual literature review in section \ref{sec:litrev}; section \ref{sec:method} details specification of the GCM-t model and derives its estimator; section \ref{sec:behav} illustrates advantages of adopting choice models with t-distributed error kernels in practice; section \ref{sec:Monte} presents a comprehensive Monte Carlo study and highlights the benefits of GCM-t over GCM-N; section \ref{sec:Empirical} validates the simulation-based findings in the empirical study; and, finally, section \ref{sec:conclusion} concludes and discusses avenues of future research. 

\section{Literature Review}\label{sec:litrev}
Regarding flexibility in modeling limited dependent variables, there is an extensive literature covering semi-parametric error distributions for binary- and multinomial-response models. However, we restrict our discussion to flexible, parsimonious parametric error specifications, and highlight research gaps that this study addresses. 

To overcome constraints from the assumption of symmetric and thin-tailed error kernels in logit and probit models, existing research offers several alternative error specifications with one or two additional shape parameters. In statistical terms, whereas an error kernel with one additional parameter (e.g., t-distribution and skew normal) allows for the trade-off between skewness and kurtosis, two additional parameters (e.g., skew t-distribution) accommodate several choices of both skewness and kurtosis. 

The scobit model derived from a Burr-10 distribution \citep{nagler1994scobit}, the robit model which considers a t-distribution with estimable DOF \citep{liu2004robit}, and the skew-probit model which assumes a skew-normal distribution with estimable skew parameter \citep{bazan2010framework} are among well-known binary response models that use error kernels with one additional parameter. Error kernels with two additional parameters, namely skewed t-distribution \citep{kim2007flexible} and the generalized Tukey lambda (GTL) family of distributions \citep{vijverberg2016pregibit} have also been applied for binary outcomes.

Some of these binary-response models with a flexible error structure have been extended to the multinomial dependent variables. \cite{castillo2008closed} and \cite{fosgerau2009discrete} derived a multinomial choice model considering a Weibull distribution on the error term in a specification that is also known as the weibit model. \cite{li2011multinomial} proposed a generalized method to construct asymmetric multinomial choice models for a family of error distributions with heteroskedastic variance, which also nests the weibit and logit models. Recently, \cite{nakayama2015unified} derived a unified multinomial choice model using the q-generalized-extreme-value (q-GEV) distribution with an estimable shape parameter and applied their model to transportation network assignment problems. \cite{brathwaite2018asymmetric} identified that all these flexible multinomial models impose restrictions on the magnitude and/or sign of the index function. To address this concern, Brathwaite and Walker proposed a generalized link function that eliminates the need for such restrictions. 

Note that all the multinomial choice models overviewed above have tractable closed-form choice probability expressions but suffer from a major limitation -- the joint modeling of multiple types of dependent variables (e.g., continuous, ordinal, and count) and the inclusion of spatial and social dependencies in these models are computationally intractable, if not impossible, due to an increase in the dimensionality of integration \citep[see][for a discussion on the curse of dimensionality in choice models]{guevara2009estimating}. 

Error kernels with skew-normal, t, or skew-t distributions have appropriate statistical properties for such joint structural modeling, but the use of these flexible distributions in multinomial choice models is cumbersome due to open-form choice probability expressions as well as noted inference issues.\footnote{Models with skew-normal and skew-t error kernel can encounter inference problems due to eventual singularity of the Fisher information matrix (when direct parameterization is used) and violation of asymptotic theory for centered parameterization \citep[see][for a discussion on these issues]{pewsey2000problems,azzalini2013maximum}.} 

The skew-normal distribution has been used at a few instances in the mixed MNP model but not to specify the error kernel, rather to model unobserved preference heterogeneity \citep[random parameter choice models as in][]{bhat2012new}. Since random taste variations and error term heterogeneity are confounded  \citep{brownstone1998forecasting}, the mixed MNP specification can be viewed as an indirect utility with a non-random index function and a skew-normal error term. \cite{bhat2015introducing} and \cite{bhat2017spatial} have also used the skew-normal distribution to account for non-normality in latent constructs within an integrated choice and latent variable model \citep{ben2002integration} and in the error kernel of an ordered response model, respectively. 

However, we are not aware of the application of t- or skew-t-distributed error kernels in multinomial choice models, which are commonly used in binary, linear mixed or multilevel, and censored linear regressions  \citep{pinheiro2001efficient,liu2004robit,koenker2009parametric,marchenko2012heckman,wang2018extending} due to their ability to model heavy-tailed distributions (Figure \ref{fig:1}). This study particularly contributes to the literature by first illustrating the statistical and behavioral implications of using a heavy-tailed error kernel in multinomial-response models and proposing the first multinomial choice model with the t-distributed error kernel.\footnote{The skew-t kernel could additionally account for asymmetry of the error distribution, but we do not consider it because inference in such models would have similar issues as we encounter in models with skew-normal kernel \citep{azzalini2013maximum}.}

\section{Methodology}\label{sec:method}
In this section, we first discuss separate models for multiple-continuous and multiple-discrete responses with t-distributed error kernels. We then combine these models to derive a generalized continuous-multinomial response model with a t-distributed error kernel (GCM-t) and outline steps for implementation of its full-information maximum likelihood estimator.\footnote{Since multinomial robit (MNR) model is a special case of the GCM-t model, we do not explicitly discuss its estimator.} In the proposed model formulation, we do not consider network, social, or spatial effects and therefore the utility of an individual is independent of other individuals in the sample. Thus, without loss of generality, we derive the model and estimation procedure for a single individual. 
   
\subsection{Continuous variable model}
Consider a standard linear regression setup: $
    y_{h} = \bm{\gamma_{h}}^{\bm{\top}} \bm{X}_h + \xi_{h},$ 
    where $h$ is the index of a continuous outcome $h = \{1,2,\dots,H\}$, $y_h$ and $\xi_h$ are the corresponding dependent variable and a t-distributed error term with DOF $\delta$, and $\bm{\gamma}_{h}$ and $\bm{X}_h$ are respectively $(s \times 1)$ vectors of coefficients and exogenous regressors. 
    
    Rewriting the regression equation in matrix form leads to: 
$\bm{y} =  \mbox{diag}(\bm{\gamma} \bm{X}^{\bm{\top}}) + \bm{\xi},$
where $(\bm{y})_{H \times 1} = [y_1,y_2, \dots, y_H]^{\top}$ is a vector of continuous outcomes  and $(\bm{\xi})_{H \times 1} = [{\xi}_1,{\xi}_2, \dots, {\xi}_H]^{\top}$ is a t-distributed error vector with DOF $\delta$, and $(\bm{\gamma})_{H \times s} = \left[\bm{\gamma_{1}},\bm{\gamma_{2}},\dots, \bm{\gamma_{H}}\right]^{\top}$ and $(\bm{X})_{H \times s} = \left[\bm{X_{1}},\bm{X_{2}},\dots, \bm{X_{H}}\right]^{\top}$ are matrices of coefficients and exogenous variables. Note that $\bm{y} \sim \mbox{MVT}_{H}\left[\mbox{diag}(\bm{\gamma} \bm{X}^{\bm{\top}}), \bm{\Xi}, \delta \right]$ where $\bm{\Xi}$ is the variance-covariance matrix of $\bm{\xi}$. 

We consider the same DOF across all elements of the error vector because the exact distribution of linear or non-linear combinations of two t-distributed random variables with  arbitrary DOF values is not known \citep{ahsanullah2014normal}. \citet{jones2002dependent} allows marginal distributions to have an arbitrary DOF in the case of a bivariate t-distributed random variable, but its extension to the multivariate case is not straightforward.        

\subsection{Choice model}
Let $i$ be the index for a nominal outcome $i \in \{1,2,\dots,I\}$, and $k$ be the index of alternatives in each nominal outcome $k \in \{1,2,\dots,i_K\}$. Then, we can write the indirect utility of alternative  $k$ in the $i^{th}$ nominal variable as $U_{ik} = \bm{\beta}_{ik}^{\bm{\top}} \bm{z}_{ik} + \epsilon_{ik},$ where $\bm{z}_{ik}$ and $\bm{\beta}_{ik}$ are $(g \times 1)$ vectors of exogenous variables and coefficients, and $\epsilon_{ik}$ is a t-distributed error term with DOF $\delta$.    

If we define the total number of alternatives $I_K = \sum \limits_{i=1}^I i_K$ , the indirect utility vector $(\bm{U})_{I_K \times 1} = \left[ \bm{U}_1, \bm{U}_2, \dots, \bm{U}_{I}\right]^{\bm{\top}}$ where $\bm{U}_{i} = \left[U_{i1},U_{i2},\dots, U_{ii_K}\right]$, the coefficient matrix $(\bm{\beta})_{I_K \times g} = \left[\bm{\beta}_{11}, \bm{\beta}_{12},\dots,\bm{\beta}_{11_{K}},\dots,\bm{\beta}_{II_K}\right]^{\top} $, the exogenuous variable matrix $(\bm{z})_{I_K \times g} = \left[\bm{z}_{11}, \bm{z}_{12},\dots,\bm{z}_{11_{K}},\dots,\bm{z}_{II_K}\right]^{\top}$, and the error vector $(\bm{\epsilon})_{I_K \times 1} = \left[ \bm{\epsilon}_1, \bm{\epsilon}_2, \dots, \bm{\epsilon}_{I}\right]^{\bm{\top}}$ where  $\bm{\epsilon}_{i} = \left[\epsilon_{i1},\epsilon_{i2},\dots, \epsilon_{ii_K}\right]$, we can write the distribution of the indirect utility as $\bm{U} \sim \mbox{MVT}_{I_K}\left[\mbox{diag}(\bm{\beta} \bm{z}^{\bm{\top}}), \bm{\Lambda},\delta \right]$ where $\bm{\Lambda}$ is the variance-covariance matrix of $\bm{\epsilon}$.

Since only differences in utility matter, the difference of error terms $(\overline{\bm{\epsilon}})$ is identifiable after fixing the scale of utility. In fact, we normalize the top diagonal element of the covariance matrix of error differences to 1 to fix scale of utility. We create a \textit{transformation matrix} $(\bm{D})$ to convert the normalized variance-covariance of error differences $(\overline{\bm{\Lambda}}_{(I_K-I) \times (I_K-I)})$ into the undifferenced error variance-covariance matrix $(\bm{\Lambda}_{I_K \times I_K})$ using $\bm{\Lambda} =\bm{D} \overline{\bm{\Lambda}} \bm{D}^{\bm{\top}}$. We provide details of creating the transformation matrix $(\bm{D})$ and an illustration of this operatoor in appendix \ref{A:D} \citep[cf.][]{bhat2012new}. The indirect utility can thus be written as $\bm{U} \sim \mbox{MVT}_{I_K}\left[\mbox{diag}(\bm{\beta} \bm{z}^{\bm{\top}}), \bm{D} \overline{\bm{\Lambda}} \bm{D}^{\bm{\top}},\delta \right]$.\footnote{The matrix  $\overline{\bm{\Lambda}}$ can be block-diagonal and still the dependencies across alternatives and nominal variables are parsimoniously generated by a single DOF parameter.} 

\subsection{Joint Model Specification}
In order to write a joint model of the continuous and nominal variable, we define $\bm{YU} = \left[\begin{array}{c}
\bm{y} \\ \bm{U} \end{array} \right]$. Thus, the distribution of $\bm{YU} \sim  \mbox{MVT}_{H+I_K}\left[ \bm{B},\bm{\Sigma},\delta \right]$ where $\bm{B} = \left[ \begin{array}{c} \mbox{diag}(\bm{\gamma} \bm{X}^{\bm{\top}}) \\ \mbox{diag}(\bm{\beta} \bm{z}^{\bm{\top}}) \end{array}  \right] $ and $\bm{\Sigma} = \left[\begin{array}{c c} \bm{\Xi} & \text{Cov}(\bm{\xi},\bm{\epsilon})\\  \text{Cov}(\bm{\epsilon}, \bm{\xi}) & \bm{\Lambda} \end{array} \right]$. If $\overline{\bm{\Sigma}}$ is the normalized (up to scale) covariance matrix of the joint differenced error $\left[\begin{array}{c} \bm{\xi}\\ \overline{\bm{\epsilon}} \end{array} \right]$, which is identified, the undifferenced full variance-covariance matrix $(\bm{\Sigma})$ can be obtained from $\overline{\bm{\Sigma}}$ using the modified transformation matrix $\bm{D}_m$ as follows: $\bm{\Sigma} =\bm{D}_m \overline{\bm{\Sigma}} \bm{D}_m^{\bm{\top}}$. Appendix \ref{A:DM} provides the details of creating $\bm{D}_m$, together with an example.  

\subsection{Joint Model Estimation}
Similar to MNP estimation, we work with utility differences using the chosen alternative as base. To perform this operation, we construct  the \textit{utility difference operator} $\bm{M}$ of size $(H+I_K-I) \times (H+I_K)$ using the algorithm given in appendix \ref{A:M}. We transform the original mean and the variance-covariance matrix using $\bm{M}$, and thus derive the distribution of the joint variable $\bm{\widetilde{YU}}$ in utility-differences $(\bm{\overline{U}})$ space. We obtain $\bm{\widetilde{YU}} \sim \mbox{MVT}_{H+I_K-I}\left(\bm{\widetilde{B}}, \bm{\widetilde{\Sigma}},\delta \right)$, where $\bm{\widetilde{B}} = \bm{M}\bm{B}$ and $\bm{\widetilde{\Sigma}} = \bm{M}\bm{\Sigma}\bm{M}^{\bm{\top}}$.    

Consider the partition of $\bm{\widetilde{B}}$ and $\bm{\widetilde{\Sigma}}$ into the continuous and choice (discrete) model (in utility differences) as follows: $\bm{\widetilde{B}} = \left[\begin{array}{c}
\bm{\widetilde{B}}_{\bm{y}} \\ \bm{\widetilde{B}}_{\bm{\overline{U}}} \end{array} \right]_{(H+I_K-I) \times 1}$ and $\bm{\widetilde{\Sigma}} = \left[\begin{array}{c c}
\bm{\widetilde{\Sigma}}_{\bm{y}} &  \bm{\widetilde{\Sigma}}_{\bm{y},\bm{\overline{U}}}\\ \bm{\widetilde{\Sigma}}_{\bm{\overline{U}}, \bm{y}} & \bm{\widetilde{\Sigma}}_{\bm{\overline{U}}} \end{array} \right]_{(H+I_K-I) \times (H+I_K-I)}$. \\

The conditional distribution of the utility difference vector is also t-distributed \citep{ding2016conditional}:   
\begin{equation}\label{eq:Udiff}
\begin{split}
& \bm{\overline{U}}| \bm{y}  \sim \mbox{MVT}_{I_K-I} \left(\overleftrightarrow{\bm{B}}_{\bm{\overline{U}}},\overleftrightarrow{\bm{\Sigma}}_{\bm{\overline{U}}}, \overleftrightarrow{\delta}\right), \\
\end{split}    
\end{equation}

where
\begin{equation*}
\begin{split}
& \overleftrightarrow{\bm{B}}_{\bm{\overline{U}}}  = \bm{\widetilde{B}}_{\bm{\overline{U}}} + \bm{\widetilde{\Sigma}}_{\bm{y},\bm{\overline{U}}}^{\bm{\top}} \left(\bm{\widetilde{\Sigma}}_{\bm{y}}\right)^{-1} (\bm{y} - \bm{\widetilde{B}}_{\bm{y}}) \\
& \overleftrightarrow{\bm{\Sigma}}_{\bm{\overline{U}}}  = \left[ \frac{\delta + \alpha}{\delta + H}\right]\left(\bm{\widetilde{\Sigma}}_{\bm{\overline{U}}} - \bm{\widetilde{\Sigma}}_{\bm{y},\bm{\overline{U}}}^{\bm{\top}} \left(\bm{\widetilde{\Sigma}}_{\bm{y}}\right)^{-1} \bm{\widetilde{\Sigma}}_{\bm{y},\bm{\overline{U}}}\right) 
\end{split}    
\end{equation*}
\begin{equation*}
\begin{split}
 \overleftrightarrow{\delta} &= \delta + H \\
 \alpha  &= (\bm{y} - \bm{\widetilde{B}}_{\bm{y}})^{\bm{\top}} \left(\bm{\widetilde{\Sigma}}_{\bm{y}}\right)^{-1}  (\bm{y} - \bm{\widetilde{B}}_{\bm{y}})
\end{split}    
\end{equation*}

Thus, the joint likelihood can be written as : 
\begin{equation}\label{eq:LL}
\begin{split}
    \mathcal{L}(\bm{\theta}) = & \mbox{Pr}(\bm{y})\mbox{Pr}(1_k = 1_m, 2_k = 2_m, \dots, I_k = I_m|\bm{y})\\
    = & f_{H}\left(\bm{y}|\bm{\widetilde{B}}_{\bm{y}},\bm{\widetilde{\Sigma}}_{\bm{y}},\delta \right) \left[\int \limits_{-\infty}^{\overleftrightarrow{\bm{B}}_{\bm{\overline{U}}}} f_{(I_K-I)}\left(r | \overleftrightarrow{\bm{B}}_{\bm{\overline{U}}},\overleftrightarrow{\bm{\Sigma}}_{\bm{\overline{U}}}, \overleftrightarrow{\delta} \right)d\bm{r}\right]
\end{split}    
\end{equation}
where $i_m$ is the chosen alternative corresponding to the $i^{th}$ nominal variable, $f_{H}$ is the probability density function of the $H$-variate t-distribution, and $\bm{\theta}$ is a vector of identified parameters $\{\bm{\gamma}, \bm{\beta}, \mbox{vec}(\bm{\overline{\Sigma}}), \delta\}$.\footnote{Note that $\bm{\overline{\Sigma}}$, the normalized variance-covariance matrix of the joint differenced error, is identified and other matrices ($\bm{\Sigma}$ and $\bm{\widetilde{\Sigma}}$) are derived from it using $\bm{D}_{m}$ and $\bm{M}$ matrices. Moreover, $\mbox{vec}(\bm{\overline{\Sigma}})$ vectorizes the unique element of a matrix $\bm{\overline{\Sigma}}$.} 

In order to ensure positive definiteness of the normalized covariance matrix of error difference ($\bm{\overline{\Sigma}}$), we work with its Cholesky decomposition in estimation. We effectively consider the normalized Cholesky factorization such that the top diagonal element of the error differenced covariance matrix of every nominal variable ($\overline{\bm{\Lambda}}_i$) is fixed to 1. Details of normalization can be found in appendix \ref{A:Chol}. 
 
The full-information maximum likelihood estimator requires maximization of the joint likelihood function. All numerical maximization routines require to evaluate the likelihood function at a given parameter vector, which we illustrate in Algorithm \ref{algo_final}. This computation involves evaluation of a  $H-$dimensional multi-variate-t-probability-density (MVTPD) function and a $(I_K - I)-$dimensional multi-variate-t-cumulative-density (MVTCD) function. The MVTCD function does not have a closed-form expression and thus requires the use of simulation-aided inference. We use a separation-of-variables (SoV) approach to compute the MVTCD function \citep{genz1999numerical}, which is detailed in section \ref{sec:MVTCD}. Similar to other simulation-based function evaluation procedures, the SOV approach also suffers from the \textit{course of dimensionality} coming from the increase in integral dimensionality. As a result, simulation-based evaluation of the function not only loses accuracy, but computation time also becomes unmanageable \citep{bhat2003simulation,craig2008new}. To reduce the dimension of this integration, we adopt the composite marginal likelihood (CML) method, which we briefly discuss in section \ref{sec:CML}.

	\begin{algorithm}[h]
	    \textbf{Input data:} $\left\{ \bm{y} , \bm{X}, \bm{z}, H, I, N \right\}$, where $N$ is the number of decision-makers; \\
		\textbf{Input parameters:} $\bm{\theta} = \left\{\bm{\gamma}, \bm{\beta}, \mbox{vec}{\left(\bm{L}_{\bm{\overline{\Sigma}}} \right)}\right\}$, where $\bm{L}_{\bm{\overline{\Sigma}}}$ is the lower triangular Cholesky matrix of $\bm{\overline{\Sigma}}$; \\
		\textbf{Step 1:} Reparametrize $\bm{L}_{\bm{\overline{\Sigma}}}$ using the procedure given in appendix \ref{A:Chol} and compute $\bm{\overline{\Sigma}}$; \\
		\textbf{Step 2:} Create \textit{modified transformation matrix} $\bm{D}_m$ using the procedure provided in appendix \ref{A:DM}; \\
		\textbf{Step 3:} Compute undifferenced error variance-covariance matrix: $\bm{\Sigma} =\bm{D}_m \overline{\bm{\Sigma}} \bm{D}_m^{\bm{\top}}$;\\
		\textbf{Step 4:} Compute mean of the joint dependent variable $\bm{YU}$ for the sample: $\bm{B} = \left[ \begin{array}{c} \mbox{diag}(\bm{\gamma} \bm{X}^{\bm{\top}}) \\ \mbox{diag}(\bm{\beta} \bm{z}^{\bm{\top}}) \end{array}  \right]$ ; \\
		\textbf{Step 5:} \\
		\SetAlgoLined
        \For{($n$ in 1 to $N$)}{
        Construct utility difference generator $\bm{M}$ using the algorithm \ref{algo_M} in appendix \ref{A:M};\\
        Compute mean $\bm{\widetilde{B}} = \bm{M}\bm{B}$ and covariance matrix $\bm{\widetilde{\Sigma}} = \bm{M}\bm{\Sigma}\bm{M}^{\bm{\top}}$ in the utility difference space;\\
        Obtain the conditional distribution of the utility difference $(\bm{\overline{U}} | \bm{y})$ using equation \ref{eq:Udiff};\\
		Compute the likelihood for the decision-maker $n$ using equation \ref{eq:LL}: \\
		\begin{itemize}
		\itemsep-.3em 
		    \item Compute PDF for the continuous variable:  $\mathcal{L}_{cont}^{n} = f_{H}\left(\bm{y}|\bm{\widetilde{B}}_{\bm{y}},\bm{\widetilde{\Sigma}}_{\bm{y}},\delta \right)$ ; 
		    \item Use CML approach (section \ref{sec:CML}) to decompose the CDF: $\mathcal{L}_{nom}^{n} = \int \limits_{-\infty}^{\overleftrightarrow{\bm{B}}_{\bm{\overline{U}}}} f_{(I_K-I)}\left(r | \overleftrightarrow{\bm{B}}_{\bm{\overline{U}}},\overleftrightarrow{\bm{\Sigma}}_{\bm{\overline{U}}}, \overleftrightarrow{\delta} \right)d\bm{r}$ ;
		    \item Compute the decomposed CDF using MVTCD function simulator (section \ref{sec:MVTCD});
		\end{itemize}
		}
        \textbf{Step 6:} Compute	the sample loglikelihood: $ \mathcal{LL} = \sum \limits_{n=1}^N \log \left[\mathcal{L}_{cont}^{n} \cdot \mathcal{L}_{nom}^{n}\right] $;
		\caption{An algorithm to compute the loglikelihood of GCM-t model} 	\label{algo_final}
	\end{algorithm}

\subsubsection{Separation-of-Variables Approach to evaluate MVTCD Function} \label{sec:MVTCD}
The underlined concept of this simulator is similar to the GHK simulator for the MVNCD function \citep{genz1992numerical,hajivassiliou1996simulation}. In the SOV approach, a $p-$dimensional integral is decomposed into $(p-1)$ unidimensional integrals using the Cholesky decomposition of the covariance matrix. These independent unidimensional integrals are sequentially evaluated based on the realization of all previous integrals. The approach presented below is adopted from \cite{genz1999numerical}. 

We present this approach for computing the area under the curve\footnote{The area under the probability density function between $\bm{a}$ and $\bm{b}$ is the same as the CDF evaluated at point $\bm{b}$ when $\bm{a}$ is $-\bm{\infty}$.} of a probability density function of the p-dimensional t-distributed random variable ($\bm{r} \sim \mbox{MVT}_{p}(\bm{0},\bm{\Omega},\delta)$)\footnote{We set location parameter to $\bm{0}$ without loss of generality because if $\bm{g} \sim \mbox{MVT}_{p}(\bm{\mu},\bm{\Omega},\delta)$ and $\bm{r} \sim \mbox{MVT}_{p}(\bm{0},\bm{\Omega},\delta)$, then  $\bm{g} = \bm{r} + \bm{\mu}$.} between $\bm{a}$ and $\bm{b}$:

\begin{equation}\label{eq:MVTCD}
T_{p}(\bm{a},\bm{b},\bm{\Omega},\delta) = \int\limits_{\bm{a}}^{\bm{b}} f(\bm{r}| \bm{\Omega}, \delta) d\bm{r} = \frac{\Gamma\left( \frac{\delta + p}{2}\right)}{\Gamma\left( \frac{\delta}{2}\right)(\pi \delta)^{\frac{p}{2}} |\bm{\Omega}|^{\frac{1}{2}}} \int\limits_{\bm{a}}^{\bm{b}} \left(1 + \frac{\bm{r}^{\bm{\top}}\bm{\Omega}^{-1}\bm{r}}{\delta} \right)^{-\frac{(\delta + p)}{2}} d\bm{r}
\end{equation}

Next, $\bm{\Omega} = \bm{L}_{\bm{\Omega}}\bm{L}_{\bm{\Omega}}^{\bm{\top}}$, where $ \bm{L}_{\bm{\Omega}}$ is the lower triangular Cholesky factor. Then, by change of variable, $\bm{r} = \bm{L}_{\bm{\Omega}}\bm{w}$ and $\bm{r}^{\bm{\top}}\bm{\Omega}^{-1}\bm{r} = \bm{w}^{\bm{\top}}\bm{w}$, $d\bm{r} = |\bm{L}_{\bm{\Omega}}|d\bm{w}$, $d\bm{r} = |\bm{\Omega}|^{\frac{1}{2}}$. Let $\kappa_{\delta}^{p} = \frac{\Gamma\left( \frac{\delta + p}{2}\right)}{\Gamma\left( \frac{\delta}{2}\right)(\pi \delta)^{\frac{p}{2}}}$ such that equation \ref{eq:MVTCD} can be rewritten as: 

\begin{equation}\label{eq:MVTCD:M1}
    T_{p}(\bm{a},\bm{b},\bm{\Omega},\delta) = \kappa_{\delta}^{p} \int \limits_{\bm{a} \leq \bm{L}_{\bm{\Omega}}\bm{w} \leq \bm{b}}  \left(1 + \frac{\bm{w}^{\bm{\top}}\bm{w}}{\delta} \right)^{-\frac{(\delta + p)}{2}} d\bm{w}  
\end{equation}
where $\bm{a} \leq \bm{L}_{\bm{\Omega}}\bm{w} \leq \bm{b}$ is the same as 
$\left[\begin{array}{c}
a_{1}\\
a_{2}\\
\vdots \\
a_{p}
\end{array}\right] \leq
\left[\begin{array}{c c c c}
L_{11} &     0  & 0 & 0\\
L_{21} & L_{22} & 0 & 0\\
\vdots & \vdots & \ddots & 0\\
L_{p1} & L_{p2} & L_{p3} & L_{pp}
\end{array}\right] \left[\begin{array}{c}
w_{1}\\
w_{2}\\
\vdots \\
w_{p}
\end{array}\right]\leq
\left[\begin{array}{c}
b_{1}\\
b_{2}\\
\vdots \\
b_{p}
\end{array}\right]$
\\

Let $\overline{a}_{i} = \frac{\left(a_{i} - \sum \limits_{j=1}^{i-1} L_{ij}w_{i} \right)}{L_{ii}}$ and $\overline{b}_{i} = \frac{\left(b_{i} - \sum \limits_{j=1}^{i-1} L_{ij}w_{i} \right)}{L_{ii}}$. Also note that 
\begin{equation}
\left(1+\frac{\bm{w}^{\bm{\top}}\bm{w}}{\delta}\right) = \left(1 + \frac{w_{1}^2}{\delta}\right) \left(1 + \frac{w_{2}^2}{\delta + w_{1}^2}\right) \dots \left(1 + \frac{w_{p}^2}{\delta + \sum \limits_{j=1}^{p-1}w_{j}^2}\right) 
\end{equation}

Thus, equation \ref{eq:MVTCD:M1} can be rewritten as : 

\begin{equation}\label{eq:MVTCD:M2}
    T_{p}(\bm{a},\bm{b},\bm{\Omega},\delta) = \kappa_{\delta}^{p} \int\limits_{\overline{a}_{1}}^{\overline{b}_{1}}\left(1 + \frac{w_{1}^2}{\delta} \right)^{-\frac{(\delta + p)}{2}}
    \int\limits_{\overline{a}_{2}}^{\overline{b}_{2}}\left(1 + \frac{w_{2}^2}{\delta + w_{1}^2}\right)^{-\frac{(\delta + p)}{2}} \dots
    \int\limits_{\overline{a}_{p}}^{\overline{b}_{p}}\left(1 + \frac{w_{p}^2}{\delta + \sum \limits_{j=1}^{p-1}w_{j}^2}\right)^{-\frac{(\delta + p)}{2}}
    d\bm{w}  
\end{equation}

Consider $w_{i} = u_i \sqrt{\frac{\delta + \sum \limits_{j=1}^{i-1}w_{j}^2}{\delta + i -1}}$. We can rewrite equation \ref{eq:MVTCD:M2} by substituting $w_{i}$ as follows: 

\begin{equation}\label{eq:MVTCD:M3}
\begin{split}
    T_{p}(\bm{a},\bm{b},\bm{\Omega},\delta) & = \kappa_{\delta}^{p} \sqrt{\left( \frac{\delta}{\delta+1}\frac{\delta}{\delta+2} \dots \frac{\delta}{\delta+p-1} \right)} \dots\\
    & \int\limits_{\hat{a}_{1}}^{\hat{b}_{1}}\left(1 + \frac{u_{1}^2}{\delta} \right)^{-\frac{(\delta + 1)}{2}}
    \int\limits_{\hat{a}_{2}}^{\hat{b}_{2}}\left(1 + \frac{u_{2}^2}{\delta + 2-1}\right)^{-\frac{(\delta + 2)}{2}} \dots
    \int\limits_{\hat{a}_{p}}^{\hat{b}_{p}}\left(1 + \frac{u_{p}^2}{\delta + p - 1}\right)^{-\frac{(\delta + p)}{2}}
    d\bm{u}  
\end{split}
\end{equation}

where $\hat{a}_{i} = \overline{a}_{i}  \sqrt{\frac{\delta + i -1}{\delta + \sum \limits_{j=1}^{i-1}w_{j}^2}}$ and $\hat{b}_{i} = \overline{b}_{i}  \sqrt{\frac{\delta + i -1}{\delta + \sum \limits_{j=1}^{i-1}w_{j}^2}}$. Further, equation \ref{eq:MVTCD:M3} can be rewritten as  

\begin{equation}\label{eq:MVTCD:M4}
\begin{split}
    T_{p}(\bm{a},\bm{b},\bm{\Omega},\delta) &=  \left[ \kappa_{\delta + 1 -1}^{1} \int\limits_{\hat{a}_{1}}^{\hat{b}_{1}}\left(1 + \frac{u_{1}^2}{\delta} \right)^{-\frac{(\delta + 1)}{2}}du_{1} \right]
    \left[\kappa_{\delta + 2 -1}^{1}\int\limits_{\hat{a}_{2}}^{\hat{b}_{2}}\left(1 + \frac{u_{2}^2}{\delta + 2-1}\right)^{-\frac{(\delta + 2)}{2}} du_{1} \right] \\ 
    & \dots \left[\kappa_{\delta + p -1}^{1} \int\limits_{\hat{a}_{p}}^{\hat{b}_{p}}\left(1 + \frac{u_{p}^2}{\delta + p - 1}\right)^{-\frac{(\delta + p)}{2}} du_{p} \right]
\end{split}
\end{equation}

The derivation of equation \ref{eq:MVTCD:M4} from equation \ref{eq:MVTCD:M2} is illustrated for a bivariate t-cumulative distribution function $(p=2)$ in appendix \ref{A:MVTCD}. Next, we substitute $u_{i} = t_{\delta + i - 1}^{-1}(z_{i})$ where $ t_{\delta + i - 1}(z_{i}) = \kappa_{\delta +i-1}^{1} \int \limits_{-\infty}^{z_i}\left(1 + \frac{s^2}{\delta + i -1} \right)^{-\frac{\delta+i}{2}}ds$ is the cumulative density function (CDF) of the univeriate t-distribution with DOF $\delta+i-1$, and thus $dz_i = k_{\delta+i-1}^{1}\left[ 1 + \frac{u_{i}^2}{\delta+i-1}\right]^{-\frac{\delta+i}{2}} du_{i}$. 

Finally,  equation \ref{eq:MVTCD:M4} can thus be rewritten as:

\begin{equation}\label{eq:MVTCD:M5}
    T_{p}(\bm{a},\bm{b},\bm{\Omega},\delta) =   \int\limits_{d_1}^{e_1}\int\limits_{d_2}^{e_2}\int\limits_{d_3}^{e_3}\dots\int\limits_{d_p}^{e_p}dz_{1}dz_{2} dz_{3}\dots dz_{p}
\end{equation}
where $d_i$ and $e_i$ are CDF of t-distributed random variable with DOF $\delta + i -1$ at points $\hat{a}_{i}$ and $\hat{b}_{i}$, respectively. After the change of variables $z_i = d_i + \phi_i (e_i-d_i)$, equation \ref{eq:MVTCD:M5} becomes: 

\begin{equation}\label{eq:MVTCD:M6}
    T_{p}(\bm{a},\bm{b},\bm{\Omega},\delta) =  (e_1-d_1)\int\limits_{0}^{1}(e_2-d_2)\dots \int\limits_{0}^{1}(e_p-d_p)\int\limits_{0}^{1} d\bm{\phi} =  \underbrace{\int\limits_{0}^{1}\int\limits_{0}^{1}\int\limits_{0}^{1}\dots\int\limits_{0}^{1}}_{(p-1)}f(\bm{\phi})d\bm{\phi}
\end{equation}
which is an integral of $f(\bm{\phi}) = \prod\limits_{i=1}^{p}(e_i-d_i)$ over the $(p-1)$-dimensional unit hyper-cube. This integral can be evaluated using different Quasi-Monte-Carlo or randomized lattice rule methods.  

\subsubsection{Composite Marginal Likelihood Approach} \label{sec:CML}
Computation of the joint likelihood function in equation \ref{eq:LL} requires the evaluation of a high-dimensional integral. The dimension of this integral grows with the number of alternatives per nominal variable and also with the number of nominal variables. For example, if there are ten nominal variables, each with six alternatives, the integral would be fifty-dimensional.        

Rather than directly evaluating such high-dimensional integrals, we use the CML approach (also known as paired-likelihood approach) for simplification. The CML method breaks down the joint likelihood function into multiple pairs, decreasing the dimension of the integral. More specifically, if choices made by an individual across all nominal variables is an event, this event is represented as pairwise observations in CML: 

\begin{equation}\label{eq:CML}
\mathcal{L}_{CML}(\bm{\theta}) = f_{H}\left(\bm{y}|\bm{\widetilde{B}}_{\bm{y}},\bm{\widetilde{\Sigma}}_{\bm{y}},\delta \right) \left(\prod \limits_{i=1}^{I-1} \prod \limits_{j=i+1}^{I} \mbox{Pr}(i_k = i_m, j_k = j_m |\bm{y}) \right)
\end{equation}

where $i_m$ represents the chosen alternative for the $i^{th}$  nominal variable. In the CML expression above, the first term is the same as the MVTPD function, but the second term corresponds to the pairing between nominal variables with the highest dimension of integration being equal to  $2[max(i_K \forall i)]$.  Thus, by employing the CML approach, the dimension of integration in the above example can be reduced to ten from fifty. Of course, the CML approach is only applicable if there are three or more nominal variables. 

A comprehensive discussion on the CML approach is outside the scope of this paper. Readers can refer to \cite{varin2005note} and \cite{varin2011overview} for a detailed discussion on the CML approach and \cite{bhat2014composite} for its derivation and application in the context of discrete choice models. Apart from well-established asymptotic properties \citep{bhat2014composite}, Bhat and co-authors have tested finite sample properties of CML by applying it to complex econometric models and have observed satisfactory results \citep{bhat2012new,bhat2015introducing,bhat2017spatial}. 

\section{Implications of using GCM-t in practice}\label{sec:behav}

\subsection{Class imbalance} \label{sec:classimb}
Class imbalance, a very high market share of few alternatives relative to others, is often encountered in choice modeling applications such as residential location choice, travel mode choice, and credit card ownership. We take an example of binary-response data to illustrate the importance of using the robit link in datasets with class imbalance. Consider a scenario where a commuter chooses an alternative between car ($C$) and bicycle ($B$). We use a data generating process with a much higher share of car than bicycle, which is reflected in the index functions: $V_C = 0.5HI - 0.2S$ and $V_B = -1-0.9HI+0.5S$, where $HI$ and $S$ are indicators for high income and student commuters, respectively.

In these class imbalance situations, accurate prediction of choices is challenging. A good model should ideally predict a higher probability of choosing bicycle (the alternative with a very low market share) when a commuter actually chooses bicycle. We compare predicted choice probabilities for all four demographic groups under normally-distributed and t-distributed error kernels with varying DOFs  (Table \ref{tab:1}). The predicted probabilities of choosing bicycle by a high-income non-student commuter using probit and robit (with 0.1 DOF) links are 0.01 and 0.38, respectively. These values are 0.38 and 0.46 for a low-income student commuter. Clearly, improvement in the predicted probability of bicycle using robit is higher when the difference between the index function values of the alternatives is higher.\footnote{The difference between the index function values of alternatives is the highest for a high-income non-student commuter and is the lowest for a low-income student commuter.}

In sum, the fat-tailed nature of the t-distributed kernel is able to better predict the probability of choosing bicycle for all demographic groups, but the extent of the improvement over probit is higher if the difference in index values lies more toward the tails of the error distribution. Thus, t-distributed error kernels increase the likelihood of predicting correct travel mode assignment in such imbalanced datasets. 

\vspace{5mm}
\centerline{\textbf{[Insert Table \ref{tab:1} here]}}
\vspace{5mm}

\subsection{Behavioral implications}
Intuitively, decision rules can be viewed as a mapping from attributes of alternatives to observed choices, which is governed by a latent construct. The latent construct includes an index function (or deterministic utility) and an idiosyncratic error term. Adoption of different decision rules manifests into different latent constructs and, thus, different mappings. Most of the previous studies have modeled different decision rules by modifying the index function. In fact, it turns out that changing the linear index function to non-linear functions in the latent construct of the fully-compensatory RUM-based model can translate it into a model with non-compensatory decision rules \citep{swait2001non, elrod2004new, martinez2009constrained}.\footnote{Whereas the decision-maker is assumed to trade all attributes of alternatives and to choose an alternative with the maximum indirect utility in the compensatory RUM framework, non-compensatory rules allow the decision-maker to choose or reject an alternative based on the value of even a single attribute \citep{schoemaker2013experiments}.} 

\begin{figure}[htbp]
\centering
\setlength{\lineskip}{\medskipamount}
\subcaptionbox{Probability density function. \label{fig:1a}}{\includegraphics[width=0.8\textwidth]{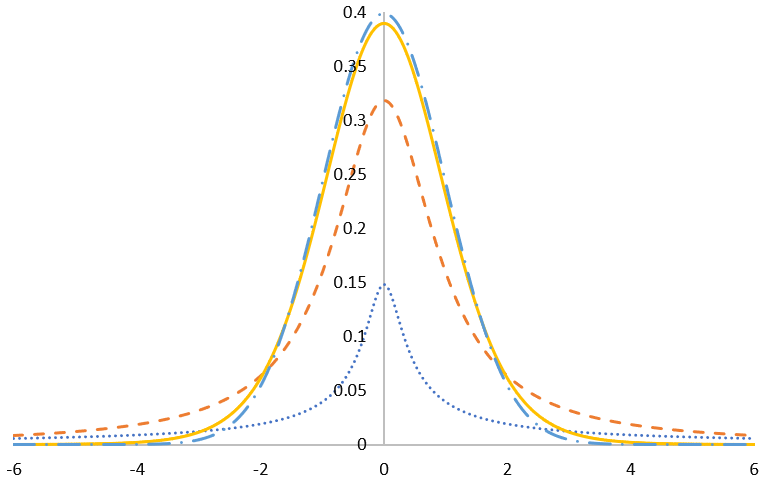}}
\subcaptionbox{Cumulative density function.\label{fig:1b}}{\includegraphics[width=0.7\textwidth]{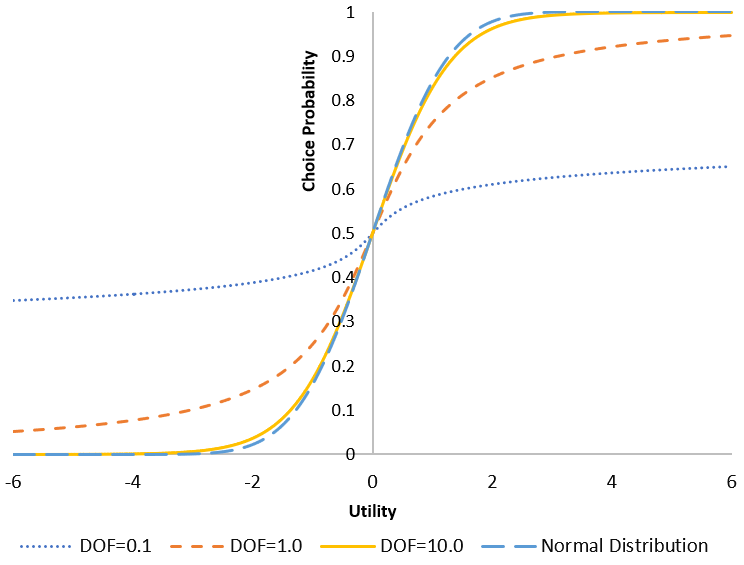}}
\caption{Comparison of t- and normally-distributed random variables.} \label{fig:1}
\end{figure}

We argue that a variation in DOF of the t-distributed error kernel across decision-makers provides a flexible way to model ``decision uncertainty" -- a degree of (un)certainty that decision-makers hold in their choices relative to the variation in the indirect utility of any alternative -- without modifying the index function in the latent construct. We illustrate this choice behavior using a plot of the choice probability vs. the utility at different DOF in a binary choice scenario (see Figure \ref{fig:1b}). Since the relationship between the choice probability and utility is steeper for the respondents with a higher DOF, choice of these decision-makers is more sensitive to a variation in utility, but in a narrower range. More specifically, \textit{decision uncertainty} of the respondents with the DOF of 10 is much larger as compared to those with the DOF of 0.1.  

In discrete choice experiments, \textit{decision uncertainty} can depend on the familiarity and experience of the decision-maker about the situation encountered during the experiment, among many other factors. Decision-makers belonging to a specific demographic group can be just more certain about their choices than others. Such choice behaviors cannot be modeled using standard logit and probit links, even when accounting for preference heterogeneity. Also, neglecting \textit{decision uncertainty} results into underestimation of welfare measures \citep{dekker2016decision}. A few studies have quantified the \textit{decision uncertainty} by asking follow-up questions after each choice task and incorporating these self-reported responses as explanatory variables or in other structural forms \citep{lundhede2009handling,olsen2011tough,beck2013consistently,borger2016fast}. 

What we propose is to capture individual-specific \textit{decision uncertainty} by parameterizing the DOF of the error kernel as a function of characeteristics of the decision maker. Since the DOF of each individual is obtained as a byproduct of estimation, GCM-t implicitly captures decision-uncertainty behavior without imposing additional econometric structure and also reduces the cognitive burden of respondents by avoiding the need of asking additional questions. 

\vspace{5mm}
\centerline{\textbf{[Insert Figure \ref{fig:1} here]}}
\vspace{5mm}

\section{Monte Carlo study and results} \label{sec:Monte}
Since this is the first study to use a t-distributed error kernel in multinomial response models, we conduct a simulation study to numerically assess the statistical properties (e.g., recovery of true parameters) of the maximum likelihood estimator of the GCM-t model. In another simulation study, we compare the performances of GCM-t and GCM-N models under thick- and thin-tailed error distributions and highlight the consequences of misspecified error kernels.

The context of our simulation study is joint modeling of the commute distance and residential location choice, i.e. integrated land-use transportation modeling. For residential location, we consider a nominal variable with five population-density-based alternatives: 0 – 99, 100-499, 500 – 1499, 1500 – 1999 and 2000 or more households/square mile. The considered joint model is shown in equation \ref{eq:DGP1}.  

To generate all three exogenous indicators, we take a draw from a standard uniform distribution. If the sampled value is higher than 0.5, the indicator takes a value 1; otherwise, it takes a value 0. That is, for a certain household if the sampled value is 0.64, 0.32, and 0.75, then the generated household is a high-income household with no children, but with a bachelor's degree holder. All the assumed parameter values and their directions are intuitive and consistent with the literature. For example, a high-income household with children prefers to live in a low-density area \citep{paleti2013integrated}. Similarly, assuming that the commute distance precedes the choice of residential location \citep{clark2003does,rashidi2012behavioral}, the likelihood of living in high-density areas decreases with the increase in the commute distance. 

Equation \ref{eq:error} shows the considered covariance matrix ($\overline{\bm{\Sigma}}$) of the joint-differenced error, which is normalized up to scale. After taking a draw from this covariance matrix, response variables are generated. Since the commute distance has a structural relationship with the choice of residential location, it is first obtained using the continuous variable model and is then used as an explanatory variable in the choice model to determine the residential location of a household. 

\begin{equation}\label{eq:DGP1}
\begin{split}
\bm{Y}\bm{U} & = \left[\begin{array}{c}
 \bm{y} \\ \hline \bm{U} \end{array} \right] = \left[\begin{array}{c}
    \text{Commute distance}\\ \hline
    \text{Density 2000+}\\
    \text{Density 0-99}\\
    \text{Density 100-499}\\
    \text{Density 500-1499}\\
    \text{Density 1500-2000}\\
\end{array}\right] \\
& = \left[\begin{array}{ccccc}
    1.00 & 0.50 & 0.75 & -0.50 & 0.00\\
    0.00 & 0.00 & 0.00 & 0.00 & 0.00 \\
   -1.50 & 1.00 & 0.90 & 0.00 & 1.00 \\
   -1.30 & 0.90 & 0.80 & 0.00 & 0.90 \\
   -1.20 & 0.80 & 0.70 & 0.00 & 0.80 \\
   -1.00 & 0.70 & 0.60 & 0.00 & 0.70\\
\end{array}\right] \left[\begin{array}{c}
    \text{Constant}  \\
    \text{High-income household}  \\
    \text{Household with children}  \\
    \text{Household with a bachelor's degree holder}  \\
    \text{Commute distance}  \\
\end{array}\right]  
\end{split}
\end{equation}

\begin{equation} \label{eq:error}
\overline{\bm{\Sigma}} = \left[\begin{array}{c|cccc}
    \textbf{1.50} & \textbf{0.30} & \textbf{0.40} & \textbf{0.60} & \textbf{0.50}\\ \hline
    \textbf{0.30} & 1.00 & 0.50 & 0.50 & 0.50\\
    \textbf{0.40} & 0.50 & \textbf{1.10} & 0.50 & 0.50 \\
    \textbf{0.60} & 0.50 & 0.50 & \textbf{1.20} & 0.50\\
    \textbf{0.50} & 0.50 & 0.50 & 0.50 & \textbf{1.30} \\
\end{array}\right]
\end{equation}

We initially estimated all elements of the covariance matrix, but such flexible specification resulted in convergence issues -- a few elements of the matrix converged to values near zero, leading to a non-positive-definite covariance matrix. This empirical identification concern is commonly encountered, forcing researchers to adopt a diagonal error-covariance matrix. Such concerns are generally not reported, but we are aware of only handful of studies with a non-diagonal error-covariance structure in multinomial choice models with open-form expressions of choice probabilities \citep[e.g.,][]{paleti2013integrated,bhat2015new}. 

In the simulation study, we estimate diagonal elements of the error-covariance matrix of the choice model and the elements representing the correlation between continuous and multinomial parts of the joint model. The estimated elements of the covariance matrix are in bold in equation \ref{eq:error}.\footnote{The fixed elements of the covariance matrix do not retain their true values during optimization due to two reasons. First, we work in the Cholesky space to ensure the positive definitiveness of the variance-covariance matrix. The non-active elements of the variance-covariance matrix are the function of active Cholesky elements which change in optimization iterations. Second, the Cholesky factor is adjusted to normalize the top diagonal element of the variance-covariance matrix of the nominal variable. This is standard practice in probit models.} We note that the diagonal error-covariance in a choice model with a t-distributed error kernel does not translate into independence across alternatives as long as the common DOF is finite.

\subsection{Statistical properties of GCM-t estimator}
To test the statistical properties of the GCM-t model, we consider the above DGP with DOF($\delta$) 1 (DGP-I) and 12 (DGP-II). Whereas a DOF value of 1 represents a distribution far from normal with flat probability curves and long thick tails, a value of 12 mimics the normal error kernel with steep probability curves and thin tails. In both DGPs, we take a sufficiently large sample of 3000 individuals to circumvent the effect of sample size on parameter recovery. Further, we generate 150 resamples for each DGP to ensure that statistical properties are not affected by the choice of the number of repetitions. To evaluate the MVTCD function, we use 200 Halton draws in the earlier discussed separation-of-variables approach. We have also tested sensitivity of parameter estimates and standard errors by increasing the number of draws from 200 to 500 for a few resamples, but have not observed any improvement in results.\footnote{200 Halton draws have proven to be sufficient to estimate up to eight-dimensional integrals in the GHK simulator for the MNP estimation \citep{patil2017simulation}.} We compute the following performance measures across resamples: \\

\noindent \textbf{Mean estimated value (MEV)}: Average value of the estimated parameter across all resamples.\\
\noindent \textbf{Mean absolute bias (MAB)}: Average bias (|true value-MEV|) of the parameter estimates.\\
\noindent \textbf{Absolute percentage bias (APB)}: |MAB/true value|$\times$100.\\ 
\noindent \textbf{Finite sample standard error (FSSE)}: Standard deviation of the parameter estimates across all resamples. \\
\noindent \textbf{Asymptotic standard error (ASE)}: Average of standard error values across all resamples, which are obtained using the sandwich estimator. For the sufficient estimator, FSSE and ASE values are close to each other.\\
\noindent \textbf{Coverage probability (CP)}: Proportion of times the 95\% confidence interval contains the true value. \\
\noindent \textbf{Power of the test}: Proportion of times the null hypothesis is rejected (i.e., |t-statistic| is greater than 1.96). \\
A lower APB, a ratio of ASE and FSSE closer to 1, higher CP, and higher power are desirable for a better statistical performance of the estimator \citep{koehler2009assessment}. \\

The resulting performance measures for DGP-I and DGP-II are summarized in Tables \ref{tab:2} and \ref{tab:3}, respectively. We first highlight important insights related to parameter recovery (i.e., bias). The mean APB values of the parameter vector ($\bm{\gamma}$) in the continuous variable model are 1.54\% and 0.46\% for DGP-I and DGP-II, respectively. These results not only indicate excellent recovery of $\bm{\gamma}$ vector, but also suggest improvement in its recovery with the increase in the DOF. However, parameters $(\bm{\beta})$ associated with the nominal response model appear to be rather difficult to recover accurately irrespective of the DOF. The mean APB values of the $\bm{\beta}$ vector are 14.79\% and 16.90\% for DGP-I and DGP-II, respectively. These values for the covariance matrix ($\overline{\bm{\Sigma}}$) are 14.16\% and 11.37\%, respectively, suggesting that the recovery of the error-covariance matrix is better for the DGP with a higher DOF value. The bias in parameter estimates is substantially higher than the ones obtained when the specification is kept the same but the normally-distributed error replaces the t-distributed error kernel in the DGP and in the estimation. More specifically, the mean APB values of $\bm{\beta}$ and $\overline{\bm{\Sigma}}$ in the corresponding GCM-N model are 4.88\% and 7.30\%, respectively.\footnote{We have also conducted a Monte Carlo study for the corresponding GCM-N model. We used the GHK simulator with 200 Halton to evaluate the MVNCD function. The detailed results are available upon request.} This comparison suggests that allowing for flexibility in the parametric distribution of the error kernel beyond mean and variance parameters might lead to a deterioration in the recovery of model parameters. \cite{bhat2012new} also observed a higher bias in the parameter estimates of the mixed multinomial choice model with a skew-normal distribution than those of the corresponding mixed MNP model. The recovery of the DOF parameter ($\delta$) in both cases is good, with APB values of 8.24\% and 12.37\%, respectively. 

We now discuss performance measures related to model inference -- ratio of asymptotic standard error and finite sample standard error (ASE/FSSE), coverage probability (CP), and power. ASE/FSEE values of all parameters $\{\bm{\gamma} ,\bm{\beta}, \overline{\bm{\Sigma}}, \delta \}$ are close to 1 for both DGPs, suggesting that the proposed estimator is sufficient. However, consistent with the findings of the parameter recovery, mean values of these ratios are much closer to 1 for DGP-II $\{0.99, 0.98, 0.95, 0.98\}$ as compared to those of the DGP-I $\{1.06, 0.81, 0.76, 1.10\}$. 
Whereas the mean CP values of the $\bm{\gamma}$ vector are close to 0.95 for both DGPs, these values are much lower for other parameters. Particularly, the mean CP value of $\bm{\beta}$ vector is 0.78 for DGPs with DOF 1 and 12, and these numbers are 0.83 and 0.90 for the error-covariance matrix $\overline{\bm{\Sigma}}$.\footnote{CP of the DOF parameter ($\delta$) in DGP-I is 0.43 (see table \ref{tab:2}). Even if the estimated DOF is close to the true DOF (low APB), such low CP occurs due to low standard errors.} Similarly, as expected, the power value of all elements of the $\bm{\gamma}$ vector is 1 for both DGPs. The estimator also provides very low variance in power values across the vector $\bm{\beta}$ with the mean values of 0.97 and 0.98 for both DGPs. However, the power variance across error-covariance matrix ($\overline{\bm{\Sigma}}$) slightly increases with the increase in DOF from 1 to 12, but the mean power value decreases from 0.80 to 0.72. Importantly, the power value of all diagonal elements of $\overline{\bm{\Sigma}}$ is 1 across both DGPs, but it is relatively lower for the off-diagonal elements.

\vspace{5mm}
\centerline{\textbf{[Insert Tables \ref{tab:2} and \ref{tab:3} here]}}
\vspace{5mm}

\subsection{Effect of modeling heavy-tailed data with normal distribution}
We conduct another Monte Carlo study to understand the consequences of assuming a normally-distributed error kernel when the actual DGP has a heavy-tailed error distribution. To accomplish this, we adopt the same model specification as described in the first Monte Carlo study, but consider a DGP with the DOF of 2 to represent a heavy-tailed distribution. We generate 50 resamples using this DGP, and estimate the GCM-t model with the estimable DOF (correct model specification) and the GCM-t model with the fixed DOF of 300 (which is equivalent to the GCM-N model) for each resample. To benchmark this analysis, we further generate 50 resamples with a thin-tailed error distribution (DOF value of 12) and estimate GCM-t with the estimable DOF and the fixed DOF of 300. The estimation results for DGPs with DOF values of 2 and 12 are provided in Tables \ref{tab:4} and \ref{tab:5}, respectively.

We obtain several insights from this second simulation study. First, modeling thick/heavy-tailed data (DOF <= 7) with a normally-distributed error kernel can introduce a large bias in the parameter estimates. The mean APB value (across all parameters) of the GCM-t model with the fixed DOF of 300 (equivalent to GCM-N model) increases from 14.46\% to 88.94\% with the decrease in DOF of the DGP from 12 to 2. These values are 11.36\% and 11.59\% for the  GCM-t model with the correct error specification. Second, we also observe a high degree of deterioration in the goodness of fit due to error misspecification. Under the heavy-tailed DGP (DOF=2), the loglikelihood value of the GCM-t model with the DOF 2 is 505 points higher than that of the GCM model with the fixed DOF of 300 (Table \ref{tab:4}). This difference in the loglikelihood values at convergence is just around 17 points under the DGP with the DOF of 12.  

We also replicate this analysis for the true DGP with DOF=1. When we estimate GCM-t with DOF=300 (i.e., equivalent GCM-N model) on resamples of this DGP, some elements of the error-covariance matrix explode and also result into the wrong direction of the parameter estimates, even when only diagonal elements are estimated. This behavior might be a manifestation of the GCM-N model's attempt to fit the index function and then broaden the support of the normal distribution around the fitted index function to put the probability mass on tails to mimic the underlying DGP. This phenomenon can potentially explain the common convergence concerns faced by researchers in estimating the error-covariance matrix of  probit-based choice models. Thus, problems in recovering the error-covariance matrix can indicate the possibility of a misspecified error kernel (i.e. the underlying assumption of normally distributed error kernel might be incorrect, and a t-distributed error kernel may perform better).

We finally compute changes in choice probabilities to reside in neighborhoods with different density levels due to a 50\% increase in commute distance for the DGP with the DOF of 2. Table \ref{tab:rev1} shows the mean and standard deviation of the elasticity estimates across 50 resamples for true GCM-t specification and GCM-t with the fixed DOF of 300 (i.e., GCM-N). We use these estimates to test the null hypothesis that differences in the mean elasticity estimates of both models are zero. The p-values of the two-tailed t-test indicate that the null hypothesis is rejected at the 5\% significance level for all alternatives. Thus, discrepancies in parameter estimates of GCM-t and GCM-N also translate into differences in the elasticity estimates. Since GCM-t recovers true parameters much better than GCM-N, its elasticity estimates are more reliable than those of GCM-N.    

\vspace{5mm}
\centerline{\textbf{[Insert Tables \ref{tab:4}, \ref{tab:5}, and \ref{tab:rev1} here]}}
\vspace{5mm}

\section{Empirical study} \label{sec:Empirical}
We now present statistical and behavioral insights from comparison of the GCM-t and GCM-N models in an empirical setting that jointly models individual's vehicle-miles-traveled (VMT) and stated vehicle purchase preferences. We also illustrate how  decision-uncertainty behavior of different demographic groups can be captured by the GCM-t model.

\subsection{Data description}
We conducted a stated preference survey of 1542 individuals in  2018 to examine the policies related to on-street parking with charging facilities for battery electric vehicles (EV) in the city of Philadelphia, Pennsylvania. Philadelphia has substantial variation in residents’ socioeconomic attributes, driving patterns, and neighborhood characteristics. The city also has the kinds of short trips and stop-and-go traffic that are well-suited to EVs. Charging locations and neighborhood parking, however, remain a substantial barrier to the adoption of EVs. In 2017, the City Council put a moratorium on a ten-year old policy to permit on-street EV charging stations and parking spaces after installing fewer than one hundred of them throughout the city. 

The main goal of the survey was to better understand preferences for adoption of electric vehicles by Philadelphians, within the context of availability of dedicated public parking with charging stations. We thus asked survey respondents to imagine themselves in a situation where they had to buy a new vehicle, had settled on a make and model, and must make a choice about whether to buy an electric version, gasoline version, or no car replacement (opt-out option) given a set of vehicle attributes. In a discrete choice experiment (DCE), each survey participant responded to purchase preferences in eight choice situations based on varying purchase prices, operating costs, electric vehicle performance in terms of driving range, and EV parking characteristics (monthly cost for access, time to recharge the battery as a proxy for type of the charging station, average time to find a parking spot as a proxy for availability, and on/off-street location). Table \ref{tab:6} presents a sample of a choice situation.

\vspace{5mm}
\centerline{\textbf{[Insert Table \ref{tab:6} here]}}
\vspace{5mm}

The experimental shares of gasoline vehicles, electric vehicles, and the opt-out alternative in the sample are 64\%, 32\%, and 4\%, respectively. The sample summary statistics of key socio-demographic and built-environment attributes are reported in Table \ref{tab:7}. Comparing our survey sample to Census micro data of persons over 17 in households with one or more cars, our respondents are substantially more likely to be female (70\% vs 52\%), white non-Hispanic (64\% vs. 36\%), younger (39 vs. 44 years old on average), and well-educated (61\% vs. 34\% with a BA or higher). The income and housing type of respondents are generally representative of Philadelphia’s adult population with cars and those who commute to work by car. For example, 63\% live in row homes and 20\% live in multi-unit buildings. Although, we had respondents from all Philadelphia Zip Codes, respondents are disproportionately from predominantly white neighborhoods in Northeast Philadelphia. In addition to the vehicle choice experiment, respondents were also asked to report their annual household vehicle-miles-traveled (VMT). In the empirical analysis, we jointly model the household's annual VMT and choice of vehicle as a function of socio-demographic, built-environment, and alternative-specific attributes.

\vspace{5mm}
\centerline{\textbf{[Insert Table \ref{tab:7} here]}}
\vspace{5mm}

\subsection{Results and discussion}
Table \ref{tab:8} summarizes the results of GCM-t and GCM-N models, which we discuss in the next subsections. 

\vspace{5mm}
\centerline{\textbf{[Insert Table \ref{tab:8} here]}}
\vspace{5mm}

\subsubsection{Socio-demographic and built-environment factors}
In this section, we discuss the relation of individual's or household's characteristics with household's annual VMT and vehicle purchase preferences. Several demographic characteristics have a significant association with the household's VMT in both GCM-t and GCM-N models. The results indicate that married households tend to have lower VMT as compared to unmarried households, ceteris paribus. This association can be a manifestation of the way different types of households spend time on their activities. For example, whereas married (or larger) households are likely to spend more time together at home in joint activities \citep{fang2008discrete,spissu2009copula}, unmarried individuals may drive more frequently for various leisure activities (e.g., going to a club, theater, or eating out) besides commuting trips. 

As expected, a household with a full-time working individual is likely to drive more than those with part-time or non-workers \citep{lee2003determinants,mcquaid2012commuting}. Surprisingly, households in higher density Zip Codes tend to drive more than those in lower density ones. This stands in sharp contrast with findings from recent meta-analyses \citep{ewing_travel_2010,stevens_does_2017}, as well as findings from the Philadelphia region more specifically \citep{klein_philadelphia_2018}. The counter-intuitive finding may relate to our sample excluding non-drivers and drawing disproportionately from higher-educated, white, women drivers in relatively dense neighborhoods of Northeast Philadelphia that are somewhat far from Philadelphia's major employment centers.

We now discuss the factors determining the vehicle purchase preferences of Philadelphia residents. We discuss the effect of socio-demographic attributes, followed by built-environment variables. First, married households with children are less inclined toward the purchase of EVs as compared to single individuals, possibly due to the compact size of  most current electric cars. Second, households with highly educated individuals (Masters or post-graduate degree) are more likely to purchase electric vehicles as compared to less educated households, perhaps as a sign of highly responsible behavior toward the planet. The literature also confirms that highly educated individuals exhibit a higher degree of environmental consciousness in terms of recycling, purchase of organic food, and awareness about factors contributing to global warming \citep{fisher2012demographic}. Third, the number of driving license holders has a positive association with the likelihood of purchasing an EV. This is because households with more drivers can better perceive the long-term benefits of buying fuel-efficient EVs as they are likely to have higher VMT. Fourth, as expected, households with a higher number of vehicles are less likely to purchase EVs. This is perhaps because car-lovers generally have higher income and may view compact design and lower range of EVs as major barriers to adoption \citep{jakobsson2016multi}. Fifth, households who already own a hybrid EV also exhibit a higher inclination toward purchasing an EV. Sixth, male respondents are more likely to purchase EVs than females. Whereas most studies have found that females are more environmental-friendly than males, some studies have also reported males to be greener than females \citep{fisher2012demographic}. Seventh, the results indicate that Asians are more likely to purchase EVs followed by Caucasians. However, low-income groups with generally less access to opportunities (such as African-American and Hispanics) are more likely to stick to traditional gasoline vehicles. Finally, as expected, millennials are more likely to purchase EVs than baby boomers and generation-X individuals. Millennials grew up in an era of prevalent information on topics such as global warming and environmental sustainability, which perhaps made them more inclined to adopt technology aimed at benefiting the environment. 

Both built-environment (walk-score and population density of the neighborhood) factors have expected and intuitive directions of effects on vehicle purchase preferences. In terms of structural effects, households with a higher vehicle mileage are less likely to purchase EVs, possibly due to limitations in driving range or lower concern for the environment.  

\subsubsection{Willingness to pay estimates}\label{sec:WTP}
We now analyze the results related to alternative-specific explanatory variables in the choice model. The directions of effects are the same and intuitive in both GCM-t and GCM-N models. Since the magnitude of marginal-utility parameters is not directly comparable, we derive willingness to pay (WTP) for improving various EV attributes. 

\begin{figure}[htbp]
\centering
\includegraphics[width=.9\textwidth]{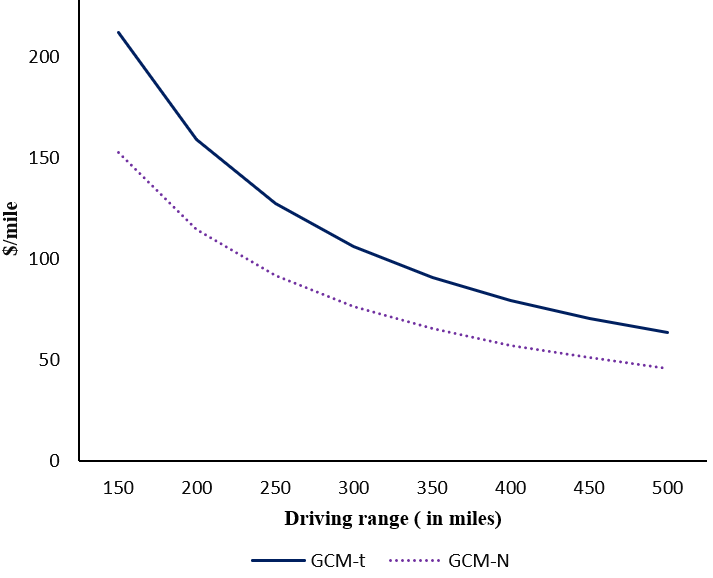}
\caption{Willingness to pay to increase the driving range of an electric vehicle by a mile} \label{fig:2}
\end{figure}

Figure \ref{fig:2} presents the maximum WTP as premium for the purchase of an electric car with a marginal improvement in driving range.\footnote{Since driving range enters the utility specification logarithmically, the WTP estimates are non-linear and decreases with the driving range.}  The results of GCM-t and GCM-N models indicate that Philadelphians are willing to pay additional \$212 (GCM-t) and \$153 (GCM-N) in purchase price, respectively, to increase driving range by one mile for a car offering 150 miles with a full battery. These WTP estimates decrease to \$64 (GCM-t) and \$46 (GCM-N) when the driving range of the car reaches 500 miles, which is close to parity with gasoline cars. Clearly, GCM-t estimates are higher than those of GCM-N, but the difference decreases with the increase in driving range.   

\vspace{5mm}
\centerline{\textbf{[Insert Figure \ref{fig:2} here]}}
\vspace{5mm}

Since monthly cost of having access to public parking with EV charging was an experimental attributes, we also estimate WTP measures for characteristics of the parking spot. For instance, we derive the WTP to reduce average search time for available EV parking. According to GCM-t and GCM-N estimates, Philadelphians are willing to add \$6.7 and \$4.6 in their monthly parking cost to reduce parking search time by a minute, respectively. This conforms to findings that on-street residential parking is frequently underpriced \citep{shoup_high_2005}. In a neighborhood where drivers spend just five minutes searching for parking on average, the average stated willingness to pay to avoid that search time is eight to eleven times higher than the price of a parking permit for a single vehicle.

WTP to reduce the EV charging time by an hour ranges from  \$96 (GCM-t) to \$60 (GCM-N). The higher WTP estimates of GCM-t are aligned with the finding of the study by \cite{dekker2016decision}, who also observed an increase in WTP for flood risk reductions after controlling for the behavioral responses to decision uncertainty in an integrated choice and latent variable model. 

Finally, because vehicle purchase and parking costs happen at different times, the annual subjective discount rate of parking cost is estimated at 10.7\% (GCM-t) and 9.3\% (GCM-N), which is slightly above market interest rates in the automotive industry.  

\subsubsection{Behavioral Insights}
We parameterize the DOF of the t-distributed error kernel as a function of demographics in GCM-t to capture  decision-uncertainty behavior of decision-makers. Table \ref{tab:9} presents the relation between the DOF and demographics in GCM-t, which is used to obtain the DOF of each respondent. These results also provide insights about decision uncertainty of different socio-demographic groups. For example, the positive relationship between married males and the DOF indicates that married males are likely to have a higher DOF, and thus are less certain about their choices as compared to unmarried females. We are not aware of any study which has measured decision-uncertainty in the context of the vehicle purchase. Therefore, beyond what observed in the current empirical analysis, we do not provide comparisons with the existing literature. 

\vspace{5mm}
\centerline{\textbf{[Insert Table \ref{tab:9} here]}}
\vspace{5mm}

We also compute elasticity estimates with respect to 1\% and 25\% reduction in the EV parking price for both models. Tables \ref{tab:10} and \ref{tab:11} present elasticity estimates for 1\% and 25\% reduction, respectively, for ten different ranges of the DOF such that each bin contains 10\% respondents of the sample.\footnote{Note that there is no concept of DOF in the GCM-N model. To facilitate the comparison between GCM-t and GCM-N models, DOF ranges are obtained from the GCM-t model and then the same set of individuals are used for respective calculations in the GCM-N models.} Whereas the difference between elasticity estimates of GCM-t and GCM-N models are not apparent for the 1\% reduction, they are quite stark for the 25\% reduction in parking price (see Table \ref{tab:11}). In general, the magnitude of elasticity estimates for gasoline and electric alternatives in the GCM-N model is higher (1.5 times, on average) than that of the corresponding GCM-t model. Specifically, this ratio is higher for respondents with lower DOF values. This supports the our earlier observation that individuals with lower DOF are more certain about their choices and are less sensitive to changes in the utility of alternatives. 

\vspace{5mm}
\centerline{\textbf{[Insert Tables \ref{tab:10} and \ref{tab:11} here]}}
\vspace{5mm}

\begin{figure}[htbp]
\centering
\setlength{\lineskip}{\medskipamount}
\subcaptionbox{Electric vehicle. \label{fig:3a}}{\includegraphics[width=.93\textwidth]{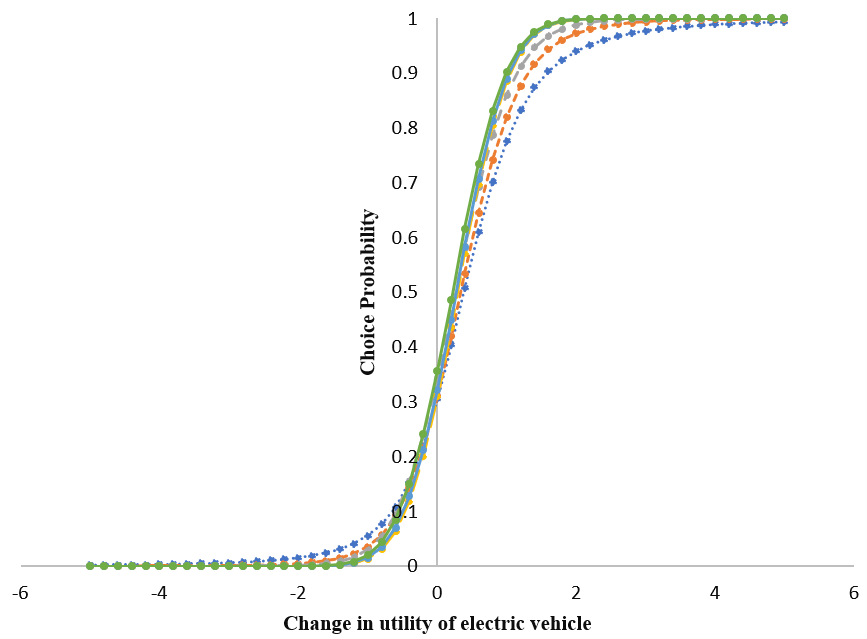}}
\subcaptionbox{Gasoline vehicle.\label{fig:3b}}{\includegraphics[width=.93\textwidth]{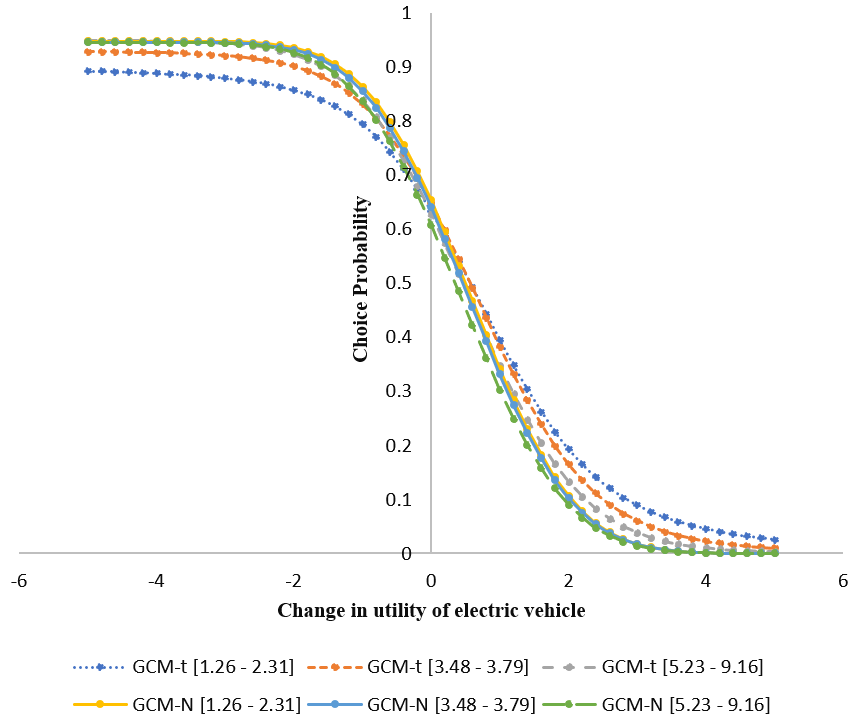}}
\caption{Predicted probabilities due to change in utility of electric vehicle.} \label{fig:3}
\end{figure}

Finally, we plot the change in the probability of choosing electric and gasoline vehicles in GCM-t and GCM-N model as a function of the change in the utility of electric vehicle in Figures \ref{fig:3a} and \ref{fig:3b}, respectively. For illustration, we only make plots for three DOF ranges. These plots offer many interesting insights. First, plots for three DOF ranges in the GCM-N model are virtually the same. Second, these change-in-probability plots become steeper with the increase in the DOF in the GCM-t model, capturing the varied decision-uncertainty behavior in the sample. Finally, curves for the lower DOFs are not asymptotic to the x-axis, demonstrating a phenomenon that it is implausible to achieve 100\% market share for any alternative in spite of being far superior to other available options. 

\vspace{5mm}
\centerline{\textbf{[Insert Figure \ref{fig:3} here]}}
\vspace{5mm}

\subsubsection{Prediction performance}
The dataset used in this case study is a good example of class-imbalance as the opt-out alternative has a very small sample share. This is an appropriate scenario to compare the predicted probability of the chosen alternative in GCM-t and GCM-N model. Table \ref{tab:12} presents the ratio of GCM-t and GCM-N choice probabilities for a group of respondents with a varying range of DOF. 

The probability ratios for two alternatives (Gasoline and Electric vehicle which have 96\% sample share in total) are close to 1 across all DOF ranges. However, the average ratio of 1.15 for the opt-out option (4\% sample share) confirms our earlier observation that GCM-t model is indeed better than a GCM-N model in modeling such imbalanced datasets. These ratios for the opt-out option are significantly higher than 1 in lower ranges of the DOF ($\leq$ 3.79). This finding is aligned with our discussion in subsection \ref{sec:classimb} on the ability of GCM-t to better predict the preference for the low-share alternatives under heavy-tailed error distributions. 

Surprisingly, GCM-N outperforms GCM-t for DOF ranges 4.16-4.40 and 5.33-9.32 with probability ratios 0.77 and 0.80, respectively. Since GCM-t can approximate GCM-N with the higher DOF, we would not expect this pattern. We speculate that such anomalies might be a manifestation of the strict constraint to use the same DOF expression across all alternatives of the choice model and the continuous response model. Nonetheless, the superior performance of GCM-N over GCM-t for certain DOF values can be leveraged by developing a latent class model with two classes where classes differ based on the distributional assumptions on the error kernel (i.e., t and normal). We discuss another flexible way to relax the ``same DOF'' assumption in section \ref{sec:conclusion}. 

We also conduct a five-fold cross-validation analysis to compare the predictive performance of GCM-t and GCM-N. We use a variant of the Brier score as a performance measure, which is a function of the predicted choice probability distribution. A lower Brier score implies better predictive performance. Table \ref{tab:rev2} shows the mean and the standard deviation of Brier scores for in-sample (70\% of total sample) and out-of-sample (30\% of total sample) predictions across five samples. In terms of the mean estimates of the Brier score, GCM-t performs better than GCM-N in in-sample predictions, but GCM-N outperforms GCM-t in out-of-sample predictions of Gasoline and opt-out alternatives. However, we cannot conclude much from this empirical study because these differences are not statistically significant due to high standard deviations. Therefore, a detailed Monte Carlo study and more empirical applications are necessary to evaluate the relative predictive performances of GCM-t and GCM-N in practice.

\vspace{5mm}
\centerline{\textbf{[Insert tables \ref{tab:12} and \ref{tab:rev2} here]}}
\vspace{5mm}

\subsubsection{Model selection}
We first compare the goodness-of-fit measures of GCM-t and GCM-N models. The log-likelihood value at convergence of GCM-t is better than the corresponding GCM-N model by 200 points (see Table \ref{tab:8}). Further, the lower Bayesian information criterion (BIC) value of GCM-t also suggests that the DGP of this empirical study is better modeled using the GCM-t model (with 39 parameters) than the GCM-N model (with 42 parameters). 

Apart from these measures, we also compare the trace of the error-covariance matrix of GCM-t and GCM-N to assess the amount of (un)explained variance in these models (see Table \ref{tab:10}). The trace values of the error-covariance matrix in GCM-t and GCM-N models are 1.06 and 3.44, respectively. This result is as expected because parameterization of the DOF in GCM-t model enables it to explain more variation in the DGP than the GCM-N model.  

\vspace{5mm}
\centerline{\textbf{[Insert Table \ref{tab:13} here]}}
\vspace{5mm}

\cite{piatek2017multinomial} did a similar effort to reduce the unexplained variance in the MNP model with latent factors. The authors split the error-covariance matrix into two parts -- the allocation (or factor loading) matrix which can be specified to model a covariance structure across alternatives, and the idiosyncratic error matrix which captures the correlation in residual. The key idea behind the allocation matrix is to allow the researcher to specify driving factors behind the choice variation beyond the observables included in the deterministic part of the utility. However, the allocation matrix is not straightforward to define in empirical studies and, beyond a mathematical inconvenience, is difficult to parametrize as a function of other variables. Thus, in contrast to splitting the error-covariance matrix, parametrization of the DOF in the GCM-t model not only offers insights about  choice behavior but also provides a new parsimonious specification to reduce the unexplained variance. 

\section{Conclusions and future work} \label{sec:conclusion}
In this study, we have proposed and applied for the first time a t-distributed error kernel within random-utility-maximization (RUM) multinomial choice models. We have not only discussed statistical advantages of using this t-distributed kernel over a normally-distributed one in class-imbalance situations, but also illustrated how the t-distributed error kernel can capture decision-uncertainty behavior (i.e., how certain the decision-makers are about the choices that they make). Furthermore, we have extended this model to a generalized continuous-multinomial response model with a t-distributed error kernel (GCM-t). Using the composite marginal likelihood method, separation-of-variables approach, and properties of the t-distribution, we have derived the full-information maximum likelihood estimator of the GCM-t model and tested its statistical properties in a Monte Carlo study. Finally, we have compared the statistical performance of GCM-t and GCM with normally-distributed error kernel (GCM-N) in a simulation study, and validated behavioral hypotheses about the advantages of GCM-t over GCM-N in an empirical study related to the adoption of electric vehicles in the city of Philadelphia.

The results of the simulation study indicate that GCM-t and GCM-N perform equally well in terms of recovering model parameters, goodness-of-fit, and model inference when the true error distribution is thin-tailed. However, GCM-t either outperforms GCM-N by a significant margin or GCM-N may not even converge in datasets with heavy-tailed error distributions. In the case study about adoption of electric vehicles, GCM-t is more accurate than GCM-N in estimating the choice of the alternative with a small sample share. Moreover, accounting for decision-uncertainty behavior in GCM-t manifests into lower elasticity estimates (relative to those of GCM-N). These lower elasticity estimates of GCM-t, as well as the evidence we found for underestimation in the GCM-N model of individual's willingness to pay to increase the driving range of electric vehicles and for reducing the parking search time for a parking spot with charging  are all relevant in the context of planning policies for broader adoption of electric vehicles.   
  
We now discuss key limitations of this study and possible research avenues to overcome them in the future. \textit{First}, unlike GCM-N, incorporating random parameters with parametric heterogeneity distributions in GCM-t is possible, but not straightforward because the resulting distribution -- the sum of t-distributed random variables or sum of t-distributed random variable and other parametric distributions -- is not of a known form. Therefore, even inclusion of parametric unobserved preference heterogeneity in the GCM-t model requires an additional layer of simulation to compute the likelihood function of the model. This necessity of adding a simulation layer would motivate researchers to rather incorporate flexible semi-parametric (instead of parametric) preference heterogeneity distributions in the GCM-t model \citep[see][for recent developments in semi-parametrics]{train2016mixed,vij2017random,bansal2018extending}. We leave this extension for future work. \textit{Second,} we considered the same degrees of freedom across all alternatives of the choice model and the continuous variable model because the exact distribution of linear or non-linear combinations of two t-distributed random variables with a arbitrary degrees-of-freedom values is not known. Extending GCM-t to a more flexible model with varying degrees of freedom across alternatives using a copula-based approach \citep{bhat2009copula} is a potential direction for future developments. This copula-based extension would allow researchers to capture decision-uncertainty behavior across alternatives, not just across decision-makers. \textit{Third}, the proposed GCM-t model cannot handle asymmetric error distributions. From an estimation standpoint, extending GCM-t model to a GCM model with a skew-t-distributed error kernel is not challenging because all relevant statistical properties are well-established in the literature \citep{azzalini2008robust}. However, practical concerns related to model inference and asymptotic properties of this new estimator require further investigation \citep{azzalini2013maximum}. \textit{Fourth}, the proposed model can be extended to accommodate both situations -- structural dependence and common unobserved errors in multinomial and continuous outcomes. For instance, the specification can be modified to simultaneously incorporate multiple causality directions by embedding a latent class framework where each class represents a causality direction. \textit{Fifth}, the current specification implicitly assigns the appropriate DOF to decision-makers, but it only captures deterministic heterogeneity in DOF. The DOF can also have variation within each demographic group. The model can be extended to capture this unobserved heterogeneity in the DOF by introducing a stochastic term in its parameterization. This generalization is straightforward but can lead to a loss in computational efficiency due to an additional layer of simulation in the estimation.

Finally, the full-information maximum likelihood estimation of GCM-t is slow because it relies on a numerical gradient in the absence of tractable expressions of the analytical gradient and Hessian matrix. The computational efficiency of this estimator can be improved by adopting tools like PyTorch or TensorFlow,  which use automatic differentiation to provide analytic gradients and Hessians without customized derivations. The possibilities of developing a Bayesian estimator \citep{kim2007flexible} or an expectation-maximization procedure \citep{liu2004robit,bansal2018minorization} can also be explored in the future to speed up the estimation of the GCM-t model. 

\section*{Acknowledgements}
 PB and RAD are thankful to the National Science Foundation CAREER Award CBET-1253475 for financially supporting methodological contributions of this research. PB is also thankful to Prof. Joan Walker and Prof. Kenneth Train for their guidance during his visit to UC Berkeley under the Exchange Scholar Program. RAD and EG also acknowledge support from a faculty research grant from the Kleinman Center for Energy Policy of the University of Pennsylvania, which supported collection of the data used as empirical case study. The authors are thankful to three anonymous reviewers for their detailed comments which significantly improved the manuscript. 

\newpage

	\bibliographystyle{agsm}
	\bibliography{references}

@article{mcfadden1973conditional,
  title={Conditional logit analysis of qualitative choice behavior},
  author={McFadden, D},
  journal={Frontiers in Econometrics},
  pages={105--142},
  year={1973},
  publisher={Academic press}
}

@article{vijverberg2016pregibit,
  title={Pregibit: a family of binary choice models},
  author={Vijverberg, Chu-Ping C and Vijverberg, Wim PM},
  journal={Empirical Economics},
  volume={50},
  number={3},
  pages={901--932},
  year={2016},
  publisher={Springer}
}

@article{fosgerau2009discrete,
  title={Discrete choice models with multiplicative error terms},
  author={Fosgerau, Mogens and Bierlaire, Michel},
  journal={Transportation Research Part B: Methodological},
  volume={43},
  number={5},
  pages={494--505},
  year={2009},
  publisher={Elsevier}
}

@article{brathwaite2018asymmetric,
  title={Asymmetric, closed-form, finite-parameter models of multinomial choice},
  author={Brathwaite, Timothy and Walker, Joan L},
  journal={Journal of Choice Modelling},
  volume={29},
  pages={78--112},
  year={2018},
  publisher={Elsevier}
}

@article{bhat2012new,
  title={A new approach to specify and estimate non-normally mixed multinomial probit models},
  author={Bhat, Chandra R and Sidharthan, Raghuprasad},
  journal={Transportation Research Part B: Methodological},
  volume={46},
  number={7},
  pages={817--833},
  year={2012},
  publisher={Elsevier}
}

@article{brownstone1998forecasting,
  title={Forecasting new product penetration with flexible substitution patterns},
  author={Brownstone, David and Train, Kenneth},
  journal={Journal of econometrics},
  volume={89},
  number={1-2},
  pages={109--129},
  year={1998},
  publisher={Elsevier}
}

@article{bhat2015introducing,
  title={Introducing non-normality of latent psychological constructs in choice modeling with an application to bicyclist route choice},
  author={Bhat, Chandra R and Dubey, Subodh K and Nagel, Kai},
  journal={Transportation Research Part B: Methodological},
  volume={78},
  pages={341--363},
  year={2015},
  publisher={Elsevier}
}

@article{liu2004robit,
  title={Robit Regression: A Simple Robust Alternative to Logistic and Probit Regression},
  author={Liu, Chuanhai},
  journal={Applied Bayesian Modeling and Causal Inference from Incomplete-Data Perspectives: An Essential Journey with Donald Rubin's Statistical Family},
  pages={227--238},
  year={2004},
  publisher={Wiley Online Library}
}

@article{pewsey2000problems,
  title={Problems of inference for Azzalini's skewnormal distribution},
  author={Pewsey, Arthur},
  journal={Journal of Applied Statistics},
  volume={27},
  number={7},
  pages={859--870},
  year={2000},
  publisher={Taylor \& Francis}
}

@article{azzalini2013maximum,
  title={Maximum penalized likelihood estimation for skew-normal and skew-t distributions},
  author={Azzalini, Adelchi and Arellano-Valle, Reinaldo B},
  journal={Journal of Statistical Planning and Inference},
  volume={143},
  number={2},
  pages={419--433},
  year={2013},
  publisher={Elsevier}
}

@article{pinheiro2001efficient,
  title={Efficient algorithms for robust estimation in linear mixed-effects models using the multivariate t distribution},
  author={Pinheiro, Jos{\'e} C and Liu, Chuanhai and Wu, Ying Nian},
  journal={Journal of Computational and Graphical Statistics},
  volume={10},
  number={2},
  pages={249--276},
  year={2001},
  publisher={Taylor \& Francis}
}

@article{genz1992numerical,
  title={Numerical computation of multivariate normal probabilities},
  author={Genz, Alan},
  journal={Journal of computational and graphical statistics},
  volume={1},
  number={2},
  pages={141--149},
  year={1992},
  publisher={Taylor \& Francis}
}

@article{wang2018extending,
  title={Extending multivariate-t linear mixed models for multiple longitudinal data with censored responses and heavy tails},
  author={Wang, Wan-Lun and Lin, Tsung-I and Lachos, Victor H},
  journal={Statistical methods in medical research},
  volume={27},
  number={1},
  pages={48--64},
  year={2018},
  publisher={SAGE Publications Sage UK: London, England}
}

@article{genz1999numerical,
  title={Numerical computation of multivariate t-probabilities with application to power calculation of multiple contrasts},
  author={Genz, Alan and Bretz, Frank},
  journal={Journal of Statistical Computation and Simulation},
  volume={63},
  number={4},
  pages={103--117},
  year={1999},
  publisher={Taylor \& Francis}
}

@article{train2016mixed,
  title={Mixed logit with a flexible mixing distribution},
  author={Train, Kenneth},
  journal={Journal of choice modelling},
  volume={19},
  pages={40--53},
  year={2016},
  publisher={Elsevier}
}

@article{ding2016conditional,
  title={On the conditional distribution of the multivariate t distribution},
  author={Ding, Peng},
  journal={The American Statistician},
  volume={70},
  number={3},
  pages={293--295},
  year={2016},
  publisher={Taylor \& Francis}
}

@article{nakayama2015unified,
  title={Unified closed-form expression of logit and weibit and its extension to a transportation network equilibrium assignment},
  author={Nakayama, Shoichiro and Chikaraishi, Makoto},
  journal={Transportation Research Part B: Methodological},
  volume={81},
  pages={672--685},
  year={2015},
  publisher={Elsevier}
}

@article{li2011multinomial,
  title={The multinomial logit model revisited: A semi-parametric approach in discrete choice analysis},
  author={Li, Baibing},
  journal={Transportation Research Part B: Methodological},
  volume={45},
  number={3},
  pages={461--473},
  year={2011},
  publisher={Elsevier}
}

@article{guevara2009estimating,
  title={Estimating random coefficient logit models with full covariance matrix: comparing performance of mixed logit and Laplace approximation methods},
  author={Guevara, Cristian Angelo and Cherchi, Elisabetta and Moreno, Matias},
  journal={Transportation Research Record},
  volume={2132},
  number={1},
  pages={87--94},
  year={2009},
  publisher={SAGE Publications Sage CA: Los Angeles, CA}
}

@article{bhat2017spatial,
  title={A spatial generalized ordered-response model with skew normal kernel error terms with an application to bicycling frequency},
  author={Bhat, Chandra R and Astroza, Sebastian and Hamdi, Amin S},
  journal={Transportation Research Part B: Methodological},
  volume={95},
  pages={126--148},
  year={2017},
  publisher={Elsevier}
}

@article{koenker2009parametric,
  title={Parametric links for binary choice models: A Fisherian--Bayesian colloquy},
  author={Koenker, Roger and Yoon, Jungmo},
  journal={Journal of Econometrics},
  volume={152},
  number={2},
  pages={120--130},
  year={2009},
  publisher={Elsevier}
}

@article{marchenko2012heckman,
  title={A Heckman selection-t model},
  author={Marchenko, Yulia V and Genton, Marc G},
  journal={Journal of the American Statistical Association},
  volume={107},
  number={497},
  pages={304--317},
  year={2012},
  publisher={Taylor \& Francis}
}

@article{varin2011overview,
  title={An overview of composite likelihood methods},
  author={Varin, Cristiano and Reid, Nancy and Firth, David},
  journal={Statistica Sinica},
  pages={5--42},
  year={2011},
  publisher={JSTOR}
}

@article{xu2011robustness,
  title={On the robustness of maximum composite likelihood estimate},
  author={Xu, Ximing and Reid, Nancy},
  journal={Journal of Statistical Planning and Inference},
  volume={141},
  number={9},
  pages={3047--3054},
  year={2011},
  publisher={Elsevier}
}

@article{hajivassiliou1996simulation,
  title={Simulation of multivariate normal rectangle probabilities and their derivatives theoretical and computational results},
  author={Hajivassiliou, Vassilis and McFadden, Daniel and Ruud, Paul},
  journal={Journal of econometrics},
  volume={72},
  number={1-2},
  pages={85--134},
  year={1996},
  publisher={Elsevier}
}

@book{de2013analysis,
  title={Analysis of mixed data: methods \& applications},
  author={De Leon, Alexander R and Chough, Keumhee Carri{\`e}re},
  year={2013},
  publisher={CRC Press}
}

@article{bhat2015new,
  title={A new generalized heterogeneous data model (GHDM) to jointly model mixed types of dependent variables},
  author={Bhat, Chandra R},
  journal={Transportation Research Part B: Methodological},
  volume={79},
  pages={50--77},
  year={2015},
  publisher={Elsevier}
}

@book{schoemaker2013experiments,
  title={Experiments on decisions under risk: The expected utility hypothesis},
  author={Schoemaker, Paul JH},
  year={2013},
  publisher={Springer Science \& Business Media}
}

@article{elrod2004new,
  title={A new integrated model of noncompensatory and compensatory decision strategies},
  author={Elrod, Terry and Johnson, Richard D and White, Joan},
  journal={Organizational Behavior and Human Decision Processes},
  volume={95},
  number={1},
  pages={1--19},
  year={2004},
  publisher={Elsevier}
}

@article{swait2001non,
  title={A non-compensatory choice model incorporating attribute cutoffs},
  author={Swait, Joffre},
  journal={Transportation Research Part B: Methodological},
  volume={35},
  number={10},
  pages={903--928},
  year={2001},
  publisher={Elsevier}
}

@article{martinez2009constrained,
  title={The constrained multinomial logit: A semi-compensatory choice model},
  author={Mart{\'\i}nez, Francisco and Aguila, Felipe and Hurtubia, Ricardo},
  journal={Transportation Research Part B: Methodological},
  volume={43},
  number={3},
  pages={365--377},
  year={2009},
  publisher={Elsevier}
}

@article{ben2002integration,
  title={Integration of choice and latent variable models},
  author={Ben-Akiva, Moshe and Walker, Joan and Bernardino, Adriana T and Gopinath, Dinesh A and Morikawa, Taka and Polydoropoulou, Amalia},
  journal={Perpetual motion: Travel behaviour research opportunities and application challenges},
  pages={431--470},
  year={2002},
  publisher={Elsevier Science}
}

@incollection{ahsanullah2014normal,
  title={Normal Distribution},
  author={Ahsanullah, Mohammad and Kibria, BM Golam and Shakil, Mohammad},
  booktitle={Normal and Student{\'{}}s t Distributions and Their Applications},
  pages={7--50},
  year={2014},
  publisher={Springer}
}

@article{jones2002dependent,
  title={A dependent bivariate t distribution with marginals on different degrees of freedom},
  author={Jones, MC},
  journal={Statistics \& probability letters},
  volume={56},
  number={2},
  pages={163--170},
  year={2002},
  publisher={Elsevier}
}

@article{craig2008new,
  title={A new reconstruction of multivariate normal orthant probabilities},
  author={Craig, Peter},
  journal={Journal of the Royal Statistical Society: Series B (Statistical Methodology)},
  volume={70},
  number={1},
  pages={227--243},
  year={2008},
  publisher={Wiley Online Library}
}

@article{bhat2003simulation,
  title={Simulation estimation of mixed discrete choice models using randomized and scrambled Halton sequences},
  author={Bhat, Chandra R},
  journal={Transportation Research Part B: Methodological},
  volume={37},
  number={9},
  pages={837--855},
  year={2003},
  publisher={Elsevier}
}

@article{bhat2014composite,
  title={The composite marginal likelihood ({CML}) inference approach with applications to discrete and mixed dependent variable models},
  author={Bhat, Chandra R},
  journal={Foundations and Trends{\textregistered} in Econometrics},
  volume={7},
  number={1},
  pages={1--117},
  year={2014},
  publisher={Now Publishers, Inc.}
}

@article{varin2005note,
  title={A note on composite likelihood inference and model selection},
  author={Varin, Cristiano and Vidoni, Paolo},
  journal={Biometrika},
  volume={92},
  number={3},
  pages={519--528},
  year={2005},
  publisher={Oxford University Press}
}

@article{paleti2013integrated,
  title={Integrated model of residential location, work location, vehicle ownership, and commute tour characteristics},
  author={Paleti, Rajesh and Bhat, Chandra R and Pendyala, Ram M},
  journal={Transportation Research Record},
  volume={2382},
  number={1},
  pages={162--172},
  year={2013},
  publisher={SAGE Publications Sage CA: Los Angeles, CA}
}

@article{rashidi2012behavioral,
  title={A behavioral housing search model: Two-stage hazard-based and multinomial logit approach to choice-set formation and location selection},
  author={Rashidi, Taha Hossein and Auld, Joshua and Mohammadian, Abolfazl Kouros},
  journal={Transportation Research Part A: Policy and Practice},
  volume={46},
  number={7},
  pages={1097--1107},
  year={2012},
  publisher={Elsevier}
}

@article{clark2003does,
  title={Does commuting distance matter?: Commuting tolerance and residential change},
  author={Clark, William AV and Huang, Youqin and Withers, Suzanne},
  journal={Regional Science and Urban Economics},
  volume={33},
  number={2},
  pages={199--221},
  year={2003},
  publisher={Elsevier}
}

@article{patil2017simulation,
  title={Simulation evaluation of emerging estimation techniques for multinomial probit models},
  author={Patil, Priyadarshan N and Dubey, Subodh K and Pinjari, Abdul R and Cherchi, Elisabetta and Daziano, Ricardo and Bhat, Chandra R},
  journal={Journal of choice modelling},
  volume={23},
  pages={9--20},
  year={2017},
  publisher={Elsevier}
}

@article{koehler2009assessment,
  title={On the assessment of Monte Carlo error in simulation-based statistical analyses},
  author={Koehler, Elizabeth and Brown, Elizabeth and Haneuse, Sebastien J-PA},
  journal={The American Statistician},
  volume={63},
  number={2},
  pages={155--162},
  year={2009},
  publisher={Taylor \& Francis}
}

@article{nagler1994scobit,
  title={Scobit: an alternative estimator to logit and probit},
  author={Nagler, Jonathan},
  journal={American Journal of Political Science},
  pages={230--255},
  year={1994},
  publisher={JSTOR}
}

@article{bazan2010framework,
  title={A framework for skew-probit links in binary regression},
  author={Baz{\'a}n, Jorge L and Bolfarine, Heleno and Branco, M{\'a}rcia D},
  journal={Communications in Statistics—Theory and Methods},
  volume={39},
  number={4},
  pages={678--697},
  year={2010},
  publisher={Taylor \& Francis}
}

@article{kim2007flexible,
  title={Flexible generalized t-link models for binary response data},
  author={Kim, Sungduk and Chen, Ming-Hui and Dey, Dipak K},
  journal={Biometrika},
  volume={95},
  number={1},
  pages={93--106},
  year={2007},
  publisher={Oxford University Press}
}

@article{castillo2008closed,
  title={Closed form expressions for choice probabilities in the Weibull case},
  author={Castillo, Enrique and Men{\'e}ndez, Jos{\'e} Mar{\'\i}a and Jim{\'e}nez, Pilar and Rivas, Ana},
  journal={Transportation Research Part B: Methodological},
  volume={42},
  number={4},
  pages={373--380},
  year={2008},
  publisher={Elsevier}
}

@article{fang2008discrete,
  title={A discrete--continuous model of households’ vehicle choice and usage, with an application to the effects of residential density},
  author={Fang, Hao Audrey},
  journal={Transportation Research Part B: Methodological},
  volume={42},
  number={9},
  pages={736--758},
  year={2008},
  publisher={Elsevier}
}

@article{spissu2009copula,
  title={A copula-based joint multinomial discrete--continuous model of vehicle type choice and miles of travel},
  author={Spissu, Erika and Pinjari, Abdul Rawoof and Pendyala, Ram M and Bhat, Chandra R},
  journal={Transportation},
  volume={36},
  number={4},
  pages={403--422},
  year={2009},
  publisher={Springer}
}

@article{lee2003determinants,
  title={Determinants of commuting time and distance for Seoul residents: The impact of family status on the commuting of women},
  author={Lee, Bun Song and McDonald, John F},
  journal={Urban Studies},
  volume={40},
  number={7},
  pages={1283--1302},
  year={2003},
  publisher={Sage Publications Sage UK: London, England}
}

@article{mcquaid2012commuting,
  title={Commuting times--The role of gender, children and part-time work},
  author={McQuaid, Ronald W and Chen, Tao},
  journal={Research in transportation economics},
  volume={34},
  number={1},
  pages={66--73},
  year={2012},
  publisher={Elsevier}
}

@article{fisher2012demographic,
  title={Demographic impacts on environmentally friendly purchase behaviors},
  author={Fisher, Caroline and Bashyal, Shristy and Bachman, Bonnie},
  journal={Journal of Targeting, Measurement and Analysis for Marketing},
  volume={20},
  number={3-4},
  pages={172--184},
  year={2012},
  publisher={Springer}
}

@article{jakobsson2016multi,
  title={Are multi-car households better suited for battery electric vehicles?--Driving patterns and economics in Sweden and Germany},
  author={Jakobsson, Niklas and Gnann, Till and Pl{\"o}tz, Patrick and Sprei, Frances and Karlsson, Sten},
  journal={Transportation Research Part C: Emerging Technologies},
  volume={65},
  pages={1--15},
  year={2016},
  publisher={Elsevier}
}

@article{piatek2017multinomial,
  title={A Multinomial Probit Model with Latent Factors: Identification and Interpretation without a Measurement System},
  author={Piatek, R{\'e}mi and Gensowski, Miriam},
  year={2017},
  publisher={IZA Discussion Paper}
}

@article{dekker2016decision,
  title={Decision uncertainty in multi-attribute stated preference studies},
  author={Dekker, Thijs and Hess, Stephane and Brouwer, Roy and Hofkes, Marjan},
  journal={Resource and Energy Economics},
  volume={43},
  pages={57--73},
  year={2016},
  publisher={Elsevier}
}

@article{bansal2018extending,
  title={Extending the logit-mixed logit model for a combination of random and fixed parameters},
  author={Bansal, Prateek and Daziano, Ricardo A and Achtnicht, Martin},
  journal={Journal of choice modelling},
  volume={27},
  pages={88--96},
  year={2018},
  publisher={Elsevier}
}

@article{vij2017random,
  title={Random taste heterogeneity in discrete choice models: Flexible nonparametric finite mixture distributions},
  author={Vij, Akshay and Krueger, Rico},
  journal={Transportation Research Part B: Methodological},
  volume={106},
  pages={76--101},
  year={2017},
  publisher={Elsevier}
}

@article{bhat2009copula,
  title={A copula-based approach to accommodate residential self-selection effects in travel behavior modeling},
  author={Bhat, Chandra R and Eluru, Naveen},
  journal={Transportation Research Part B: Methodological},
  volume={43},
  number={7},
  pages={749--765},
  year={2009},
  publisher={Elsevier}
}

@article{azzalini2008robust,
  title={Robust likelihood methods based on the skew-t and related distributions},
  author={Azzalini, Adelchi and Genton, Marc G},
  journal={International Statistical Review},
  volume={76},
  number={1},
  pages={106--129},
  year={2008},
  publisher={Wiley Online Library}
}

@article{lundhede2009handling,
  title={Handling respondent uncertainty in choice experiments: evaluating recoding approaches against explicit modelling of uncertainty},
  author={Lundhede, Thomas Hedemark and Olsen, S{\o}ren B{\o}ye and Jacobsen, Jette Bredahl and Thorsen, Bo Jellesmark},
  journal={Journal of Choice Modelling},
  volume={2},
  number={2},
  pages={118--147},
  year={2009},
  publisher={Elsevier}
}

@article{olsen2011tough,
  title={Tough and easy choices: testing the influence of utility difference on stated certainty-in-choice in choice experiments},
  author={Olsen, S{\o}ren B{\o}ye and Lundhede, Thomas Hedemark and Jacobsen, Jette Bredahl and Thorsen, Bo Jellesmark},
  journal={Environmental and Resource Economics},
  volume={49},
  number={4},
  pages={491--510},
  year={2011},
  publisher={Springer}
}

@article{beck2013consistently,
  title={Consistently inconsistent: The role of certainty, acceptability and scale in choice},
  author={Beck, Matthew J and Rose, John M and Hensher, David A},
  journal={Transportation Research Part E: Logistics and Transportation Review},
  volume={56},
  pages={81--93},
  year={2013},
  publisher={Elsevier}
}

@article{borger2016fast,
  title={Are fast responses more random? Testing the effect of response time on scale in an online choice experiment},
  author={B{\"o}rger, Tobias},
  journal={Environmental and Resource Economics},
  volume={65},
  number={2},
  pages={389--413},
  year={2016},
  publisher={Springer}
}

@article{ewing_travel_2010,
	title = {Travel and the built environment: a meta-analysis},
	volume = {76},
	issn = {0194-4363},
	url = {http://www.informaworld.com/10.1080/01944361003766766},
	doi = {10.1080/01944361003766766},
	number = {3},
	urldate = {2010-06-28},
	journal = {Journal of the American Planning Association},
	author = {Ewing, Reid and Cervero, Robert},
	year = {2010},
	pages = {265--294}
}

@book{shoup_high_2005,
	address = {Chicago},
	title = {The {High} {Cost} of {Free} {Parking}},
	isbn = {1-884829-98-8},
	publisher = {Planners Press, American Planning Association},
	author = {Shoup, Donald},
	year = {2005}
}

@article{stevens_does_2017,
	title = {Does compact development make people drive less?},
	volume = {83},
	url = {http://www.tandfonline.com/doi/abs/10.1080/01944363.2016.1240044},
	doi = {10.1080/01944363.2016.1240044},
	number = {1},
	urldate = {2017-04-19},
	journal = {Journal of the American Planning Association},
	author = {Stevens, Mark R.},
	year = {2017},
	pages = {7--18}
}

@article{klein_philadelphia_2018,
	title = {The {Philadelphia} story: {Age}, race, gender and changing travel trends},
	volume = {69},
	issn = {0966-6923},
	shorttitle = {The {Philadelphia} story},
	url = {http://www.sciencedirect.com/science/article/pii/S0966692317307044},
	doi = {10.1016/j.jtrangeo.2018.04.009},
	urldate = {2018-05-15},
	journal = {Journal of Transport Geography},
	author = {Klein, Nicholas J. and Guerra, Erick and Smart, Michael J.},
	month = may,
	year = {2018},
	pages = {19--25}
}

@article{bansal2018minorization,
  title={Minorization-Maximization (MM) algorithms for semiparametric logit models: Bottlenecks, extensions, and comparisons},
  author={Bansal, Prateek and Daziano, Ricardo A and Guerra, Erick},
  journal={Transportation Research Part B: Methodological},
  volume={115},
  pages={17--40},
  year={2018},
  publisher={Elsevier}
}
\newpage

\begin{appendix}
\section{Matrix transformations} 
\subsection{\texorpdfstring{Transformation matrix $(\bm{D})$ to compute $\bm{\Lambda}$ from $\overline{\bm{\Lambda}}$}{Transformation matrix}} \label{A:D}
We show below the relation between the variance-covariance matrix of the error $\left(\bm{\Lambda}_{I_K \times I_K} \right)$ and the normalized variance-covariance matrix of the error difference $\left(\bm{\overline{\Lambda}}_{(I_K-I) \times (I_K-I)}\right)$: 

\begin{equation}
    \begin{split}
\text{Variance-covariance of error difference}   \quad      \bm{\overline{\Lambda}}&= 
\left[\begin{array}{c c c c}
\bm{\overline{\Lambda}}_{1} & \bm{\overline{\Lambda}}_{1,2} & \dots & \bm{\overline{\Lambda}}_{1,I} \\
\bm{\overline{\Lambda}}_{2,1} & \bm{\overline{\Lambda}}_{2} & \dots &  \bm{\overline{\Lambda}}_{2,I} \\
\vdots & \vdots & \ddots & \vdots \\
\bm{\overline{\Lambda}}_{I,1} & \bm{\overline{\Lambda}}_{I,2} & \dots & \bm{\overline{\Lambda}}_{I} 
\end{array}\right]_{(I_K-I) \times (I_K-I)} \\
\text{where} \quad \bm{\overline{\Lambda}}_{i} &=  
\left[\begin{array}{c c c c}
1 & \overline{\Lambda}_{1,2}^{i} & \dots &  \overline{\Lambda}_{1,i_K-1}^{i} \\
\overline{\Lambda}_{2,1}^{i} & \overline{\Lambda}_{2,2}^{i} & \dots & \overline{\Lambda}_{2,i_K-1}^{i} \\
\vdots & \vdots & \ddots & \vdots \\
\overline{\Lambda}_{i_K-1,1}^{i} & \overline{\Lambda}_{i_K-1,2}^{i} & \dots & \overline{\Lambda}_{i_K-1,i_K-1}^{i} \\
\end{array}\right]_{(i_K-1) \times (i_K-1)} \\
\text{Variance-covariance of error}  \quad   \bm{\Lambda} & = 
\left[\begin{array}{c c c c}
\bm{\Lambda}_{1} & \bm{\Lambda}_{1,2} & \dots & \bm{\Lambda}_{1,I} \\
\bm{\Lambda}_{2,1} & \bm{\Lambda}_{2} & \dots &  \bm{\Lambda}_{2,I} \\
\vdots & \vdots & \ddots & \vdots \\
\bm{\Lambda}_{I,1} & \bm{\Lambda}_{I,2} & \dots & \bm{\Lambda}_{I} 
\end{array}\right]_{(I_K) \times (I_K)} \\
where \quad  \bm{\Lambda}_{i} & = 
\left[\begin{array}{c c}
0 & \bm{0}_{1,i_K-1} \\
\bm{0}_{i_K-1,1} & \bm{\overline{\Lambda}}_{i}
\end{array}\right]_{i_K \times i_K}  \\
and \quad \bm{\Lambda}_{i,j} & = 
\left[\begin{array}{c c}
0 & \bm{0}_{1,j_K-1} \\
\bm{0}_{i_K-1,1} & \bm{\overline{\Lambda}}_{i,j}
\end{array}\right]_{i_K \times j_K}
    \end{split}
\end{equation}

Note that $\bm{\Lambda} = \bm{D} \bm{\overline{\Lambda}} \bm{D}^{\bm{\top}}$, where the transformation matrix $\bm{D}$ is constructed based on algorithm \ref{algo_D}. 

	\begin{algorithm}[h]
		\textbf{Initialisation:} $\bm{D} = \bm{0}_{I_K \times (I_K-I)}$ \\ 
		\SetAlgoLined
        \For{($m$ in 1 to $I$)}{
		
     	\eIf{$m==1$}{
		$R_1 = 2$ ; \\
		$R_2 = m_K$; \\
		$C_1 = 1$ ; \\
		$C_2 = m_K - 1$ ;\\
		}{
		$R_1 = \sum \limits_{n=1}^{m-1}n_K + 2$ ; \\
		$R_2 =  \sum \limits_{n=1}^{m}n_K$; \\
		$C_1 =  \sum \limits_{n=1}^{m-1}(n_K - 1) + 1$ ; \\
		$C_2 = \sum \limits_{n=1}^{m}(n_K - 1) $ ;\\
		}
		$\bm{D}(R_1:R_2,C_1:C_2) = \bm{1}_{(m_K - 1) \times (m_K - 1)}$ ;\\
		}
		{\small \textbf{Note:} $\bm{1}_{i \times i}$ and $\bm{0}_{i \times i}$ are identity matrix and matrix of zeros, respectively, of size $i \times i$.}
		\caption{Creating $\bm{D}$ matrix} 	\label{algo_D}
	\end{algorithm}

We provide an example to illustrate the transformation from $\bm{\overline{\Lambda}}$ to $\bm{\Lambda}$ using $\bm{D}$. We consider a case of $I=2$, with $1_K = 3$ and $2_K = 4$. The $\bm{\overline{\Lambda}}_{5 \times 5}$, $\bm{\Lambda}_{7 \times 7}$, and the transformation matrix $\bm{D}_{7 \times 5}$ would be: 

\begin{equation}
    \begin{split}
    \bm{\overline{\Lambda}} &= \left[\begin{array}{c c}
    \bm{\overline{\Lambda}}_{1} & \bm{\overline{\Lambda}}_{1,2} \\
    \bm{\overline{\Lambda}}_{2,1} & \bm{\overline{\Lambda}}_{2}
    \end{array}\right] = \left[\begin{array}{c c|c c c}
        1 & \overline{\Lambda}_{1,2}^{1} & \overline{\Lambda}_{1,1}^{1,2} & \overline{\Lambda}_{1,2}^{1,2} & \overline{\Lambda}_{1,3}^{1,2} \\  
        \overline{\Lambda}_{2,1}^{1} & \overline{\Lambda}_{2,2}^{1} &  \overline{\Lambda}_{2,1}^{1,2} & \overline{\Lambda}_{2,2}^{1,2} & \overline{\Lambda}_{2,3}^{1,2} \\ \hline
        \overline{\Lambda}_{1,1}^{2,1} & \overline{\Lambda}_{1,2}^{2,1} &  1 &  \overline{\Lambda}_{1,2}^{2} & \overline{\Lambda}_{1,3}^{2} \\
        \overline{\Lambda}_{2,1}^{2,1} & \overline{\Lambda}_{2,2}^{2,1} & \overline{\Lambda}_{2,1}^{2}  &  \overline{\Lambda}_{2,2}^{2} & \overline{\Lambda}_{2,3}^{2} \\
        \overline{\Lambda}_{3,1}^{2,1} & \overline{\Lambda}_{3,2}^{2,1} & \overline{\Lambda}_{3,1}^{2}  &  \overline{\Lambda}_{3,2}^{2} & \overline{\Lambda}_{3,3}^{2} \\
        \end{array}\right]
    \\
    \bm{\Lambda}  &= \left[\begin{array}{c c}
    \bm{\Lambda}_{1} & \bm{\Lambda}_{1,2} \\
    \bm{\Lambda}_{2,1} & \bm{\Lambda}_{2}
    \end{array}\right] = = \left[\begin{array}{c c c|c c c c} 
        0 & 0 & 0 & 0 & 0 & 0 & 0 \\
        0 & 1 & \overline{\Lambda}_{1,2}^{1} & 0 & \overline{\Lambda}_{1,1}^{1,2} & 0 \overline{\Lambda}_{1,2}^{1,2} & \overline{\Lambda}_{1,3}^{1,2} \\  
        0 & \overline{\Lambda}_{2,1}^{1} & \overline{\Lambda}_{2,2}^{1} &  0 & \overline{\Lambda}_{2,1}^{1,2} & \overline{\Lambda}_{2,2}^{1,2} & \overline{\Lambda}_{2,3}^{1,2} \\ \hline
        0 & 0 & 0 & 0 & 0 & 0 & 0 \\
        0 & \overline{\Lambda}_{1,1}^{2,1} & \overline{\Lambda}_{1,2}^{2,1} & 0 & 1 &  \overline{\Lambda}_{1,2}^{2} & \overline{\Lambda}_{1,3}^{2} \\
        0 & \overline{\Lambda}_{2,1}^{2,1} & \overline{\Lambda}_{2,2}^{2,1} & 0 & \overline{\Lambda}_{2,1}^{2}  &  \overline{\Lambda}_{2,2}^{2} & \overline{\Lambda}_{2,3}^{2} \\
        0 & \overline{\Lambda}_{3,1}^{2,1} & \overline{\Lambda}_{3,2}^{2,1} & 0 & \overline{\Lambda}_{3,1}^{2}  &  \overline{\Lambda}_{3,2}^{2} & \overline{\Lambda}_{3,3}^{2} \\
        \end{array}\right] \\
    \bm{D} &= \left[\begin{array}{c c c c c}
    0 & 0 & 0 & 0 & 0 \\
    1 & 0 & 0 & 0 & 0 \\
    0 & 1 & 0 & 0 & 0 \\
    0 & 0 & 0 & 0 & 0 \\
    0 & 0 & 1 & 0 & 0 \\
    0 & 0 & 0 & 1 & 0 \\
    0 & 0 & 0 & 0 & 1 \\    
\end{array}\right]
 \end{split}
\end{equation}

\subsection{\texorpdfstring{Modified transformation matrix $(\bm{D}_m)$ to compute $\bm{\Sigma}$ from $\overline{\bm{\Sigma}}$}{Modified transformation matrix}}  \label{A:DM}
The \textit{modified transformation matrix} $(\bm{D}_m)$ can be easily computed from the \textit{transformation matrix} $(\bm{D})$ by appending an identity matrix as follows: \\
\begin{equation}
    \bm{D}_m =  \left[\begin{array}{c c}   \bm{1}_{H \times H} & \bm{0}_{H \times (I_K - I)} \\ \bm{0}_{I_K \times H} &  \bm{D} \end{array}\right]_{(H + I_K) \times (H + I_K - I)}
\end{equation}
where $H$ is the number of continuous outcomes, $\bm{1}_{H \times H}$ is an identity matrix of size $H \times H$, and $\bm{D}$ is obtained from algorithm \ref{algo_D}. We expand on the example of appendix \ref{A:D} which considers $I=2$ with $1_K = 3$ and $2_K = 4$. If $H=2$, then the \textit{modified transformation matrix} $\bm{D}_{m}$ would be:

\begin{equation}
    \begin{split}
     \bm{D}_m &= \left[\begin{array}{c c}   \bm{1}_{2 \times 2} & \bm{0}_{2 \times 5} \\ \bm{0}_{7 \times 2} &  \bm{D}_{7 \times 5} \end{array}\right]\\
     & = \left[\begin{array}{c c | c c c c c}
    1 & 0 & 0 & 0 & 0 & 0 & 0 \\
    0 & 1 & 0 & 0 & 0 & 0 & 0 \\ \hline
    0 & 0 & 0 & 0 & 0 & 0 & 0 \\
    0 & 0 & 1 & 0 & 0 & 0 & 0 \\
    0 & 0 & 0 & 1 & 0 & 0 & 0 \\
    0 & 0 & 0 & 0 & 0 & 0 & 0 \\
    0 & 0 & 0 & 0 & 1 & 0 & 0 \\
    0 & 0 & 0 & 0 & 0 & 1 & 0 \\
    0 & 0 & 0 & 0 & 0 & 0 & 1 \\    
\end{array}\right] 
 \end{split}
\end{equation}
 
\subsection{\texorpdfstring{Utility difference generator $(M)$ to compute $\bm{\widetilde{\Sigma}}$ from $\bm{\Sigma}$}{Utility difference generator}} \label{A:M}
	\begin{algorithm}[h]
		\textbf{Initialisation:} $\bm{M} = \bm{0}_{(H+I_K-I) \times (H+I_K)}$ \; 
		$\bm{M}(1:H,1:H) =\bm{1}_{H \times H}$ \;
		\SetAlgoLined
        \For{($n$ in 1 to $I$)}{
        $F_{1} = \bm{1}_{(n_K - 1) \times (n_K - 1)}$ \;
        $T_{1} = -\bm{O}_{n_K - 1}$ \;
        \eIf{$n_m ==1$}{
        $S_1 =T_{1} \sim F_{1}$ \;
        }{\eIf{$n_m ==n_K$}{
        $S_1 =F_{1} \sim T_{1}$ \;
        }{
        $S_1 =F_{1}[,1:(n_m-1)] \sim T_{1} \sim F_{1}[,n_m:(n_K-1)]$\;
        }
		}
		\eIf{$n==1$}{
		$R_1 = H + 1$ ; \\
		$R_2 = H + n_K -1$; \\
		$C_1 = H + 1$ ; \\
		$C_2 = H + n_K$ ;\\
		}{
		$R_1 = H + \left( \sum \limits_{j=1}^{n-1}(j_K - 1) \right)+ 1$ ; \\
		$R_2 = H + \left(  \sum \limits_{j=1}^{n}(j_K-1) \right)$; \\
		$C_1 = H + \left( \sum \limits_{j=1}^{n-1}j_K  \right) + 1$ ; \\
		$C_2 = H + \left( \sum \limits_{j=1}^{n}j_K  \right)$ ;\\
		}
		$\bm{M}(R_1:R_2,C_1:C_2) = S_1$\;
		}
		{\small \textbf{Note 1:} $\bm{1}_{i \times i}$ and $\bm{0}_{i \times i}$ are identity matrix and matrix of zeros, respectively, of size $i \times i$. \\
		\textbf{Note 2:} $\bm{O}_{i}$ is a column vector of ones of size $i \times 1$. $``\sim"$ implies horizontal concatenation.  \\
		\textbf{Note 3:} $n_m$ is the index of the chosen alternative for the nominal variable $n$.\\}
		\caption{Creating $\bm{M}$ matrix} 	\label{algo_M}
	\end{algorithm}

 We consider the same example of appendix \ref{A:DM} which considers $H=2$ and $I=2$ with $1_K = 3$, $1_m = 2$, $2_K = 4$, and $2_m=3$. We use algorithm \ref{algo_M} to construct $\bm{M}$ matrix for this example:

\begin{equation}
   \bm{M} = \left[\begin{array}{c c | c c c c c c c}
    1 & 0 & 0 & 0 & 0 & 0 & 0 & 0 & 0\\
    0 & 1 & 0 & 0 & 0 & 0 & 0 & 0 & 0\\ \hline
    0 & 0 & 1 & -1 & 0 & 0 & 0 & 0 & 0\\
    0 & 0 & 0 & -1 & 1 & 0 & 0 & 0 & 0\\
    0 & 0 & 0 & 0 & 0 & 1 & 0 & -1 & 0\\
    0 & 0 & 0 & 0 & 0 & 0 & 1 & -1 & 0\\
    0 & 0 & 0 & 0 & 0 & 0 & 0 & -1 & 1\\
\end{array}\right] 
\end{equation}

\subsection{\texorpdfstring{Reparametrization of the Cholesky decomposition of $\bm{\overline{\Sigma}}$}{reparametrization of the Cholesky decomposition of Sigma}}  \label{A:Chol}
Consider $\bm{L}\bm{L}^{\bm{\top}} = \bm{\overline{\Sigma}}$, where $\bm{L}$ is the lower triangular Cholesky matrix of size $(H+I_K-I) \times (H+I_K-I)$. We reparametrize all rows of the $\bm{L}$ matrix for which the diagonal element of $\bm{\overline{\Sigma}}$ is normalized to 1. We compute $a_i = \sqrt{1 + \sum \limits_{j=1}^{i-1} L_{i,j}^{2}}$ for the $i^{th}$ row and modify non-diagonal elements $L_{i,r} = \frac{L_{i,r}}{a_i} \forall r \in \{1,2,\dots,i-1\}$ and the diagonal element $L_{i,i} = \frac{1}{a_i}$.

\section{\texorpdfstring{MVTNCD illustration}{MVTNCD illustration}}  \label{A:MVTCD}
The derivation of equation \ref{eq:MVTCD:M4} from equation \ref{eq:MVTCD:M2} can be understood easily based on a transformation applied on a bivariate t-cumulative distribution function. For $p=2$ , equation \ref{eq:MVTCD:M2} after substitution can be written as (note that $w_1 = u_1$ ):
\begin{equation*}
\begin{split}
    T_{p}(\bm{a},\bm{b},\bm{\Omega},\delta)  &= \kappa_{\delta}^{2}     \int\limits_{\hat{a}_{1}}^{\hat{b}_{1}}\left(1 + \frac{u_{1}^2}{\delta} \right)^{-\frac{(\delta + 2)}{2}}
    \int\limits_{\hat{a}_{2}}^{\hat{b}_{2}}\left(1 + \frac{u_{2}^2}{\delta + 2-1}\right)^{-\frac{(\delta + 2)}{2}} d{u}_{1} \left( d{u}_{2}\sqrt{\frac{\delta + u_{1}^2}{\delta+1}}\right ) \\
    & = \kappa_{\delta}^{2} \left[ \int\limits_{\hat{a}_{1}}^{\hat{b}_{1}}\sqrt{\frac{\delta + u_{1}^2}{\delta+1}}  \left(1 + \frac{u_{1}^2}{\delta} \right)^{-\frac{(\delta + 2)}{2}}  d{u}_{1} \right] 
    \left[\int\limits_{\hat{a}_{2}}^{\hat{b}_{2}}\left(1 + \frac{u_{2}^2}{\delta + 2-1}\right)^{-\frac{(\delta + 2)}{2}}d{u}_{2} \right]\\
    & = \kappa_{\delta}^{2} \left[ \int\limits_{\hat{a}_{1}}^{\hat{b}_{1}}\sqrt{\left(\frac{\delta}{\delta+1}\right) \left(1 + \frac{u_{1}^2}{\delta} \right)}  \left(1 + \frac{u_{1}^2}{\delta} \right)^{-\frac{(\delta + 2)}{2}}  d{u}_{1} \right] 
    \left[\int\limits_{\hat{a}_{2}}^{\hat{b}_{2}}\left(1 + \frac{u_{2}^2}{\delta + 2-1}\right)^{-\frac{(\delta + 2)}{2}}d{u}_{2} \right]\\
    & = \kappa_{\delta}^{2} \sqrt{\frac{\delta}{\delta+1}}  \int\limits_{\hat{a}_{1}}^{\hat{b}_{1}}  \left(1 + \frac{u_{1}^2}{\delta} \right)^{-\frac{(\delta + 1)}{2}} 
    \int\limits_{\hat{a}_{2}}^{\hat{b}_{2}}\left(1 + \frac{u_{2}^2}{\delta + 2-1}\right)^{-\frac{(\delta + 2)}{2}}  d{u}_{1} d{u}_{2} \quad \text{(equation \ref{eq:MVTCD:M3})}\\
    & \text{Now deriving equation \ref{eq:MVTCD:M4}} \\
    & = \frac{\Gamma\left( \frac{\delta + 2}{2}\right)}{\Gamma\left( \frac{\delta}{2}\right) (\pi \delta)} \sqrt{\frac{\delta}{\delta+1}}  \int\limits_{\hat{a}_{1}}^{\hat{b}_{1}}  \left(1 + \frac{u_{1}^2}{\delta} \right)^{-\frac{(\delta + 1)}{2}} 
    \int\limits_{\hat{a}_{2}}^{\hat{b}_{2}}\left(1 + \frac{u_{2}^2}{\delta + 2-1}\right)^{-\frac{(\delta + 2)}{2}}  d{u}_{1} d{u}_{2} \\ 
    & =  \frac{\Gamma\left( \frac{\delta+1}{2}\right)\Gamma\left( \frac{\delta+1+1}{2}\right)}
    {\Gamma\left( \frac{\delta}{2}\right)(\pi \delta)^{\frac{1}{2}}\Gamma\left( \frac{\delta+1}{2}\right)(\pi (\delta+1))^{\frac{1}{2}}}  
    \int\limits_{\hat{a}_{1}}^{\hat{b}_{1}}  \left(1 + \frac{u_{1}^2}{\delta} \right)^{-\frac{(\delta + 1)}{2}} 
    \int\limits_{\hat{a}_{2}}^{\hat{b}_{2}}\left(1 + \frac{u_{2}^2}{\delta + 2-1}\right)^{-\frac{(\delta + 2)}{2}}  d{u}_{1} d{u}_{2} \\ 
    & =  \kappa_{\delta}^{1} \kappa_{\delta+1}^{1}\int\limits_{\hat{a}_{1}}^{\hat{b}_{1}}  \left(1 + \frac{u_{1}^2}{\delta} \right)^{-\frac{(\delta + 1)}{2}} 
    \int\limits_{\hat{a}_{2}}^{\hat{b}_{2}}\left(1 + \frac{u_{2}^2}{\delta + 2-1}\right)^{-\frac{(\delta + 2)}{2}}  d{u}_{1} d{u}_{2} \\
    & = \left[ \kappa_{\delta + 1 -1}^{1} \int\limits_{\hat{a}_{1}}^{\hat{b}_{1}}\left(1 + \frac{u_{1}^2}{\delta} \right)^{-\frac{(\delta + 1)}{2}}du_{1} \right]
    \left[\kappa_{\delta + 2 -1}^{1}\int\limits_{\hat{a}_{2}}^{\hat{b}_{2}}\left(1 + \frac{u_{2}^2}{\delta + 2-1}\right)^{-\frac{(\delta + 2)}{2}} du_{1} \right] \text{(equation \ref{eq:MVTCD:M4})}
\end{split}
\end{equation*}
\end{appendix}

\newpage

\centerline{\Large{\textbf{Figures}}}

\clearpage
\newpage

\centerline{\Large{\textbf{Tables}}}

\begin{table}[h!]
		\centering
		\caption{Class-imbalance example under Probit and t-distributed error kernel \vspace{-.3cm}}
		\label{tab:1}
		\begin{adjustwidth}{-1.1cm}{}
		\resizebox{1.1\textwidth}{!}{
\begin{tabular}{|l|cc|cc|cc|cc|}
\hline
\multirow{3}{*}{\textbf{Demographic profile}} & \multicolumn{6}{c|}{\textbf{t-distributed kernel}} & \multicolumn{2}{c|}{\multirow{2}{*}{\textbf{Probit kernel}}} \\ \cline{2-7}
 & \multicolumn{2}{c}{\textbf{DOF = 0.1}} & \multicolumn{2}{c}{\textbf{DOF = 0.5}} & \multicolumn{2}{c|}{\textbf{DOF = 1.0}} & \multicolumn{2}{c|}{} \\ \cline{2-9} 
 & \textbf{P(car)} & \textbf{P(bicycle)} & \textbf{P(car)} & \textbf{P(bicycle)} & \textbf{P(car)} & \textbf{P(bicycle)} & \textbf{P(car)} & \textbf{P(bicycle)} \\ \hline
High-income and non-student & 0.62 & 0.38 & 0.80 & 0.20 & 0.88 & 0.12 & 0.99 & 0.01 \\
High-income and student	& 0.60 & 0.40 & 0.76 & 0.24 & 0.83 & 0.17 & 0.96 & 0.04 \\
Low-income and non-student & 0.58 & 0.42 & 0.70 & 0.30 & 0.75 & 0.25 & 0.84 & 0.16 \\
Low-income and student & 0.54 & 0.46 & 0.58 & 0.42 & 0.59 & 0.41 & 0.62 & 0.38 \\  \hline
\end{tabular}}
\end{adjustwidth}
\end{table}

\begin{table}[h!]
		\centering
		\caption{Simulation Results of GCM-t Model for DOF-I scenario (DOF=1) \vspace{-.3cm}}
		\label{tab:2}
		\begin{adjustwidth}{0cm}{}
		\resizebox{1\textwidth}{!}{
\begin{tabular}{|cccccccccc|}
\hline
\textbf{Parameter} & \textbf{True Value} & \textbf{MEV} & \textbf{MAB} & \textbf{APB} & \textbf{FSSE} & \textbf{ASE} & \textbf{ASE/ FSSE} & \textbf{CP} & \textbf{Power} \\ \hline
\multicolumn{10}{|c|}{\textbf{Mean effect elements}} \\ \hline
$\gamma_{11}$ & 1 & 0.98 & 0.02 & 1.69 & 0.06 & 0.06 & 1.07 & 0.93 & 1 \\
$\gamma_{12}$ & 0.5 & 0.51 & 0.01 & 2.06 & 0.06 & 0.06 & 1.12 & 0.97 & 1 \\
$\gamma_{13}$ & 0.75 & 0.76 & 0.01 & 1.46 & 0.07 & 0.06 & 0.98 & 0.93 & 1 \\
$\gamma_{14}$ & -0.5 & -0.5 & 0 & 0.94 & 0.06 & 0.06 & 1.06 & 0.97 & 1 \\
$\beta_{21}$ & -1.5 & -1.74 & 0.24 & 16.23 & 0.28 & 0.16 & 0.56 & 0.47 & 1 \\
$\beta_{22}$ & 1 & 0.91 & 0.09 & 8.71 & 0.11 & 0.11 & 0.99 & 0.89 & 1 \\
$\beta_{23}$ & 0.9 & 0.81 & 0.09 & 10.35 & 0.13 & 0.12 & 0.98 & 0.9 & 1 \\
$\beta_{25}$ & 1 & 0.94 & 0.06 & 6.46 & 0.26 & 0.16 & 0.63 & 0.72 & 0.99 \\
$\beta_{31}$ & -1.3 & -1.47 & 0.17 & 13.21 & 0.26 & 0.2 & 0.77 & 0.85 & 1 \\
$\beta_{32}$ & 0.9 & 0.79 & 0.11 & 12.35 & 0.14 & 0.12 & 0.87 & 0.81 & 1 \\
$\beta_{33}$ & 0.8 & 0.68 & 0.12 & 15.58 & 0.16 & 0.13 & 0.84 & 0.79 & 1 \\
$\beta_{35}$ & 0.9 & 0.8 & 0.1 & 10.9 & 0.23 & 0.17 & 0.73 & 0.8 & 0.99 \\
$\beta_{41}$ & -1.2 & -1.36 & 0.16 & 13.49 & 0.27 & 0.2 & 0.76 & 0.79 & 1 \\
$\beta_{42}$ & 0.8 & 0.67 & 0.13 & 16.54 & 0.15 & 0.13 & 0.88 & 0.76 & 1 \\
$\beta_{43}$ & 0.7 & 0.57 & 0.13 & 18.7 & 0.19 & 0.15 & 0.81 & 0.8 & 0.93 \\
$\beta_{45}$ & 0.8 & 0.68 & 0.12 & 15.23 & 0.25 & 0.18 & 0.74 & 0.77 & 0.92 \\
$\beta_{51}$ & -1 & -1.08 & 0.08 & 7.72 & 0.22 & 0.18 & 0.83 & 0.91 & 1 \\
$\beta_{52}$ & 0.7 & 0.54 & 0.16 & 23.33 & 0.15 & 0.12 & 0.85 & 0.65 & 0.99 \\
$\beta_{53}$ & 0.6 & 0.43 & 0.17 & 27.66 & 0.17 & 0.15 & 0.84 & 0.76 & 0.83 \\
$\beta_{55}$ & 0.7 & 0.56 & 0.14 & 20.14 & 0.23 & 0.18 & 0.81 & 0.81 & 0.8 \\ \hline
\multicolumn{10}{|c|}{\textbf{Covariance matrix elements}} \\ \hline
$\overline{\Sigma}_{11}$ & 1.5 & 1.62 & 0.12 & 8.28 & 0.1 & 0.09 & 0.97 & 0.75 & 1 \\
$\overline{\Sigma}_{21}$ & 0.3 & 0.28 & 0.02 & 8.22 & 0.29 & 0.18 & 0.61 & 0.77 & 0.42 \\
$\overline{\Sigma}_{31}$ & 0.4 & 0.44 & 0.04 & 9.78 & 0.3 & 0.2 & 0.66 & 0.79 & 0.6 \\
$\overline{\Sigma}_{33}$ & 1.1 & 1.21 & 0.11 & 10.41 & 0.29 & 0.24 & 0.83 & 0.89 & 1 \\
$\overline{\Sigma}_{41}$ & 0.6 & 0.72 & 0.12 & 20.24 & 0.35 & 0.24 & 0.7 & 0.78 & 0.76 \\
$\overline{\Sigma}_{44}$ & 1.2 & 1.48 & 0.28 & 23.19 & 0.4 & 0.33 & 0.83 & 0.91 & 1 \\
$\overline{\Sigma}_{51}$ & 0.5 & 0.6 & 0.1 & 19.12 & 0.34 & 0.24 & 0.72 & 0.82 & 0.65 \\
$\overline{\Sigma}_{55}$ & 1.3 & 1.48 & 0.18 & 14.06 & 0.41 & 0.31 & 0.76 & 0.9 & 1 \\ \hline
\multicolumn{10}{|c|}{\textbf{Degree-of-freedom}} \\ \hline
$\delta$ & 1 & 1.08 & 0.08 & 8.24 & 0.03 & 0.04 & 1.1 & 0.43 & 1 \\ \hline
\textbf{Average} &  &  & \textbf{0.14} & \textbf{12.56} & \textbf{0.21} & \textbf{0.16} & \textbf{0.84} & \textbf{0.8} & \textbf{0.93} \\ \hline
\end{tabular}}
\footnotesize{\textbf{Note:} Some values may be zero due to rounding off.}\\
\footnotesize{\textbf{Acronyms:} MEV: mean estimated value, MAB: mean absolute bias, APB: absolute percentage bias, FSSE: finite sample standard error, ASE: asymptotic standard error, CP: coverage probability.}
\end{adjustwidth}
\end{table}

\begin{table}[h!]
		\centering
		\caption{Simulation Results of GCM-t Model for DOF-II scenario (DOF=12) \vspace{-.3cm}}
		\label{tab:3}
		\begin{adjustwidth}{0cm}{}
		\resizebox{1\textwidth}{!}{
\begin{tabular}{|cccccccccc|}
\hline
\textbf{Parameter} & \multicolumn{1}{l}{\textbf{True Value}} & \textbf{MEV} & \textbf{MAB} & \textbf{APB} & \multicolumn{1}{l}{\textbf{FSSE}} & \multicolumn{1}{l}{\textbf{ASE}} & \multicolumn{1}{l}{\textbf{ASE/ FSSE}} & \multicolumn{1}{l}{\textbf{CP}} & \multicolumn{1}{l|}{\textbf{Power}} \\ \hline
\multicolumn{10}{|c|}{\textbf{Mean effect elements}} \\ \hline
$\gamma_{11}$ & 1 & 1 & 0 & 0.38 & 0.04 & 0.05 & 1.09 & 0.97 & 1 \\
$\gamma_{12}$ & 0.5 & 0.5 & 0 & 0.77 & 0.05 & 0.05 & 0.99 & 0.97 & 1 \\
$\gamma_{13}$ & 0.75 & 0.75 & 0 & 0.42 & 0.05 & 0.05 & 0.94 & 0.91 & 1 \\
$\gamma_{14}$ & -0.5 & -0.5 & 0 & 0.26 & 0.05 & 0.05 & 0.92 & 0.91 & 1 \\
$\beta_{21}$ & -1.5 & -1.76 & 0.26 & 17.62 & 0.17 & 0.15 & 0.88 & 0.49 & 1 \\
$\beta_{22}$ & 1 & 0.89 & 0.11 & 11.48 & 0.11 & 0.11 & 0.96 & 0.8 & 1 \\
$\beta_{23}$ & 0.9 & 0.79 & 0.11 & 12.12 & 0.12 & 0.12 & 0.98 & 0.87 & 1 \\
$\beta_{25}$ & 1 & 0.9 & 0.1 & 10.28 & 0.18 & 0.19 & 1.08 & 0.94 & 1 \\
$\beta_{31}$ & -1.3 & -1.42 & 0.12 & 9.58 & 0.27 & 0.27 & 1.01 & 0.95 & 1 \\
$\beta_{32}$ & 0.9 & 0.76 & 0.14 & 16.03 & 0.13 & 0.13 & 1.02 & 0.77 & 1 \\
$\beta_{33}$ & 0.8 & 0.67 & 0.13 & 16.38 & 0.15 & 0.14 & 0.94 & 0.77 & 1 \\
$\beta_{35}$ & 0.9 & 0.74 & 0.16 & 17.89 & 0.18 & 0.19 & 1.06 & 0.83 & 0.99 \\
$\beta_{41}$ & -1.2 & -1.29 & 0.09 & 7.49 & 0.25 & 0.26 & 1.04 & 0.97 & 1 \\
$\beta_{42}$ & 0.8 & 0.63 & 0.17 & 21.82 & 0.13 & 0.12 & 0.92 & 0.65 & 1 \\
$\beta_{43}$ & 0.7 & 0.53 & 0.17 & 24.39 & 0.16 & 0.13 & 0.83 & 0.62 & 0.98 \\
$\beta_{45}$ & 0.8 & 0.64 & 0.16 & 20.04 & 0.19 & 0.19 & 1.04 & 0.87 & 0.93 \\
$\beta_{51}$ & -1 & -1.02 & 0.02 & 1.86 & 0.2 & 0.22 & 1.11 & 0.98 & 1 \\
$\beta_{52}$ & 0.7 & 0.5 & 0.2 & 28.48 & 0.12 & 0.11 & 0.94 & 0.52 & 1 \\
$\beta_{53}$ & 0.6 & 0.42 & 0.18 & 30.18 & 0.14 & 0.12 & 0.88 & 0.61 & 0.92 \\
$\beta_{55}$ & 0.7 & 0.53 & 0.17 & 24.78 & 0.18 & 0.19 & 1.02 & 0.79 & 0.81 \\ \hline
\multicolumn{10}{|c|}{\textbf{Covariance matrix elements}} \\ \hline
$\overline{\Sigma}_{11}$ & 1.5 & 1.52 & 0.02 & 1.22 & 0.06 & 0.06 & 1.05 & 0.97 & 1 \\
$\overline{\Sigma}_{21}$ & 0.3 & 0.17 & 0.13 & 43.74 & 0.21 & 0.19 & 0.9 & 0.85 & 0.19 \\
$\overline{\Sigma}_{31}$ & 0.4 & 0.34 & 0.06 & 15.32 & 0.23 & 0.2 & 0.87 & 0.9 & 0.41 \\
$\overline{\Sigma}_{33}$ & 1.1 & 1.1 & 0 & 0.04 & 0.41 & 0.41 & 1.01 & 0.93 & 1 \\
$\overline{\Sigma}_{41}$ & 0.6 & 0.55 & 0.05 & 8.92 & 0.29 & 0.23 & 0.81 & 0.87 & 0.67 \\
$\overline{\Sigma}_{44}$ & 1.2 & 1.21 & 0.01 & 0.53 & 0.48 & 0.48 & 1.01 & 0.9 & 1 \\
$\overline{\Sigma}_{51}$ & 0.5 & 0.43 & 0.07 & 13.6 & 0.26 & 0.22 & 0.84 & 0.89 & 0.49 \\
$\overline{\Sigma}_{55}$ & 1.3 & 1.2 & 0.1 & 7.58 & 0.4 & 0.45 & 1.1 & 0.91 & 1 \\ \hline
\multicolumn{10}{|c|}{\textbf{Degree-of-freedom}} \\ \hline
$\delta$ & 12 & 13.48 & 1.48 & 12.37 & 2.89 & 2.84 & 0.98 & 0.97 & 1 \\ \hline
\textbf{Average} &  &  & \textbf{0.18} & \textbf{12.95} & \textbf{0.28} & \textbf{0.27} & \textbf{0.97} & \textbf{0.84} & \textbf{0.91} \\ \hline
\end{tabular}}
\footnotesize{\textbf{Note:} Some values may be zero due to rounding off.}\\
\footnotesize{\textbf{Acronyms:} MEV: mean estimated value, MAB: mean absolute bias, APB: absolute percentage bias, FSSE: finite sample standard error, ASE: asymptotic standard error, CP: coverage probability.}
\end{adjustwidth}
\end{table}

\begin{table}[h!]
		\centering
		\caption{Effect of ignoring non-normality (DOF=2) \vspace{-.3cm}}
		\label{tab:4}
		\begin{adjustwidth}{1.5cm}{}
		\resizebox{.8\textwidth}{!}{
\begin{tabular}{|l|c|ccc|ccc|}
\hline
\multirow{2}{*}{\textbf{Parameter}} & \multirow{2}{*}{\textbf{True Value}} & \textbf{MEV} & \textbf{MAB} & \textbf{APB} & \textbf{MEV} & \textbf{MAB} & \textbf{APB} \\ \cline{3-8} 
 &  & \multicolumn{3}{c}{\textbf{GCM-t with DOF of 2}} & \multicolumn{3}{c|}{\textbf{GCM-t with DOF of 300}} \\ \hline
\multicolumn{8}{|c|}{\textbf{Mean effect elements}} \\ \hline
$\gamma_{11}$ & 1 & 0.98 & 0.02 & 1.53 & 0.98 & 0.02 & 1.6 \\
$\gamma_{12}$ & 0.5 & 0.51 & 0.01 & 2.12 & 0.51 & 0.01 & 2.51 \\
$\gamma_{13}$ & 0.75 & 0.76 & 0.01 & 0.99 & 0.76 & 0.01 & 1.93 \\
$\gamma_{14}$ & -0.5 & -0.5 & 0 & 0.67 & -0.51 & 0.01 & 2.48 \\
$\beta_{21}$ & -1.5 & -1.75 & 0.25 & 16.91 & -1.4 & 0.1 & 6.36 \\
$\beta_{22}$ & 1 & 0.9 & 0.1 & 10.13 & 0.66 & 0.34 & 33.79 \\
$\beta_{23}$ & 0.9 & 0.81 & 0.09 & 9.94 & 0.6 & 0.3 & 32.91 \\
$\beta_{25}$ & 1 & 0.91 & 0.09 & 8.95 & 0.65 & 0.35 & 35.18 \\
$\beta_{31}$ & -1.3 & -1.49 & 0.19 & 14.34 & -1.33 & 0.03 & 2.44 \\
$\beta_{32}$ & 0.9 & 0.76 & 0.14 & 16.1 & 0.55 & 0.35 & 38.33 \\
$\beta_{33}$ & 0.8 & 0.67 & 0.13 & 16.66 & 0.49 & 0.31 & 38.6 \\
$\beta_{35}$ & 0.9 & 0.82 & 0.08 & 9.33 & 0.6 & 0.3 & 33.77 \\
$\beta_{41}$ & -1.2 & -1.36 & 0.16 & 13.28 & -1.16 & 0.04 & 2.98 \\
$\beta_{42}$ & 0.8 & 0.64 & 0.16 & 20.45 & 0.44 & 0.36 & 45.38 \\
$\beta_{43}$ & 0.7 & 0.5 & 0.2 & 28.62 & 0.33 & 0.37 & 52.55 \\
$\beta_{45}$ & 0.8 & 0.71 & 0.09 & 10.75 & 0.5 & 0.3 & 37.95 \\
$\beta_{51}$ & -1 & -1.09 & 0.09 & 8.82 & -2.65 & 1.65 & 164.78 \\
$\beta_{52}$ & 0.7 & 0.53 & 0.17 & 24.31 & 0.34 & 0.36 & 51.03 \\
$\beta_{53}$ & 0.6 & 0.42 & 0.18 & 30.29 & 0.25 & 0.35 & 58.77 \\
$\beta_{55}$ & 0.7 & 0.55 & 0.15 & 21.58 & 0.28 & 0.42 & 60.32 \\ \hline
\multicolumn{8}{|c|}{\textbf{Covariance matrix elements}} \\ \hline
$\overline{\Sigma}_{11}$ & 1.5 & 1.57 & 0.07 & 4.42 & 4.2 & 2.7 & 180.26 \\
$\overline{\Sigma}_{21}$ & 0.3 & 0.24 & 0.06 & 21.37 & -0.37 & 0.67 & 222.38 \\
$\overline{\Sigma}_{31}$ & 0.4 & 0.32 & 0.08 & 19.77 & -0.18 & 0.58 & 145.74 \\
$\overline{\Sigma}_{33}$ & 1.1 & 1.09 & 0.01 & 0.72 & 1.54 & 0.44 & 39.58 \\
$\overline{\Sigma}_{41}$ & 0.6 & 0.55 & 0.05 & 9.13 & 0.31 & 0.29 & 48.77 \\
$\overline{\Sigma}_{44}$ & 1.2 & 1.23 & 0.03 & 2.25 & 1.39 & 0.19 & 15.44 \\
$\overline{\Sigma}_{51}$ & 0.5 & 0.51 & 0.01 & 1.49 & 1.07 & 0.57 & 113.11 \\
$\overline{\Sigma}_{55}$ & 1.3 & 1.37 & 0.07 & 5.48 & 14.58 & 13.28 & 1021.4 \\ \hline
\multicolumn{8}{|c|}{\textbf{Degree-of-freedom}} \\ \hline
$\delta$ & 2 & 2.12 & 0.12 & 5.84 &  &  &  \\
\textbf{Average} &  &  & \textbf{0.1} & \textbf{11.59} &  & \textbf{0.88} & \textbf{88.94} \\ \hline
Loglikelihood & \multicolumn{1}{l}{} & \multicolumn{3}{c}{-10417.44} & \multicolumn{3}{c|}{-10922.73} \\ \hline
\end{tabular}}
\\ \footnotesize{\textbf{Note:} Some values may be zero due to rounding off.}\\
\footnotesize{\textbf{Acronyms:} MEV: mean estimated value, MAB: mean absolute bias, APB: absolute percentage bias.}
\end{adjustwidth}
\end{table}

\begin{table}[h!]
     	\centering
		\caption{Effect of ignoring non-normality (DOF=12) \vspace{-.3cm}}
		\label{tab:5}
		\begin{adjustwidth}{1.5cm}{}
		\resizebox{.8\textwidth}{!}{
\begin{tabular}{|l|c|ccc|ccc|}
\hline
\multirow{2}{*}{\textbf{Parameter}} & \multirow{2}{*}{\textbf{True Value}} & \textbf{MEV} & \textbf{MAB} & \textbf{APB} & \textbf{MEV} & \textbf{MAB} & \textbf{APB} \\ \cline{3-8} 
 &  & \multicolumn{3}{c}{\textbf{GCM-t with DOF of 12}} & \multicolumn{3}{c|}{\textbf{GCM-t with DOF of 300}} \\ \hline
\multicolumn{8}{|c|}{\textbf{Mean effect elements}} \\ \hline
$\gamma_{11}$ & 1 & 1 & 0 & 0.39 & 1 & 0 & 0.32 \\
$\gamma_{12}$ & 0.5 & 0.49 & 0.01 & 1.46 & 0.49 & 0.01 & 1.77 \\
$\gamma_{13}$ & 0.75 & 0.75 & 0 & 0.45 & 0.75 & 0 & 0.03 \\
$\gamma_{14}$ & -0.5 & -0.5 & 0 & 0.26 & -0.5 & 0 & 0.01 \\
$\beta_{21}$ & -1.5 & -1.75 & 0.25 & 16.83 & -1.7 & 0.2 & 13.54 \\
$\beta_{22}$ & 1 & 0.89 & 0.11 & 11 & 0.85 & 0.15 & 14.99 \\
$\beta_{23}$ & 0.9 & 0.81 & 0.09 & 10.02 & 0.78 & 0.12 & 13.87 \\
$\beta_{25}$ & 1 & 0.89 & 0.11 & 10.84 & 0.85 & 0.15 & 14.84 \\
$\beta_{31}$ & -1.3 & -1.4 & 0.1 & 7.98 & -1.36 & 0.06 & 4.97 \\
$\beta_{32}$ & 0.9 & 0.77 & 0.13 & 13.98 & 0.74 & 0.16 & 18.16 \\
$\beta_{33}$ & 0.8 & 0.72 & 0.08 & 10.01 & 0.69 & 0.11 & 13.78 \\
$\beta_{35}$ & 0.9 & 0.71 & 0.19 & 20.82 & 0.68 & 0.22 & 24.95 \\
$\beta_{41}$ & -1.2 & -1.32 & 0.12 & 10.17 & -1.31 & 0.11 & 9.15 \\
$\beta_{42}$ & 0.8 & 0.63 & 0.17 & 21 & 0.6 & 0.2 & 25.02 \\
$\beta_{43}$ & 0.7 & 0.54 & 0.16 & 22.19 & 0.52 & 0.18 & 25.77 \\
$\beta_{45}$ & 0.8 & 0.64 & 0.16 & 19.57 & 0.61 & 0.19 & 24.17 \\
$\beta_{51}$ & -1 & -1.03 & 0.03 & 2.95 & -1.04 & 0.04 & 4.31 \\
$\beta_{52}$ & 0.7 & 0.52 & 0.18 & 25.83 & 0.49 & 0.21 & 29.86 \\
$\beta_{53}$ & 0.6 & 0.45 & 0.15 & 25.09 & 0.42 & 0.18 & 29.24 \\
$\beta_{55}$ & 0.7 & 0.52 & 0.18 & 26.39 & 0.49 & 0.21 & 30.27 \\\hline
\multicolumn{8}{|c|}{\textbf{Covariance matrix elements}} \\ \hline
$\overline{\Sigma}_{11}$ & 1.5 & 1.51 & 0.01 & 0.33 & 1.77 & 0.27 & 17.8 \\
$\overline{\Sigma}_{21}$ & 0.3 & 0.2 & 0.1 & 32.59 & 0.15 & 0.15 & 50.2 \\
$\overline{\Sigma}_{31}$ & 0.4 & 0.4 & 0 & 0.16 & 0.38 & 0.02 & 4.28 \\
$\overline{\Sigma}_{33}$ & 1.1 & 1.13 & 0.03 & 3.15 & 1.14 & 0.04 & 3.63 \\
$\overline{\Sigma}_{41}$ & 0.6 & 0.57 & 0.03 & 5.12 & 0.59 & 0.01 & 1.82 \\
$\overline{\Sigma}_{44}$ & 1.2 & 1.32 & 0.12 & 10 & 1.41 & 0.21 & 17.19 \\
$\overline{\Sigma}_{51}$ & 0.5 & 0.47 & 0.03 & 5.93 & 0.48 & 0.02 & 4.8 \\
$\overline{\Sigma}_{55}$ & 1.3 & 1.28 & 0.02 & 1.77 & 1.38 & 0.08 & 6.22 \\ \hline
\multicolumn{8}{|c|}{\textbf{Degree-of-freedom}} \\ \hline
$\delta$ & 12 & 13.56 & 1.56 & 13.04 &  &  &  \\
\textbf{Average} &  &  & \textbf{0.14} & \textbf{11.36} &  & \textbf{0.12} & \textbf{14.46} \\ \hline
Loglikelihood &  & \multicolumn{3}{c|}{-8977.45} & \multicolumn{3}{c|}{-8994.84} \\ \hline
\end{tabular}}
\\ \footnotesize{\textbf{Note:} Some values may be zero due to rounding off.}\\
\footnotesize{\textbf{Acronyms:} MEV: mean estimated value, MAB: mean absolute bias, APB: absolute percentage bias.}
\end{adjustwidth}
\end{table}

\begin{table}[h!]
		\centering
		\caption{Change in probability due to 50\% increase in commute distance (DGP with DOF of 2) \vspace{-.3cm}}
		\label{tab:rev1}
		\begin{adjustwidth}{-0.5cm}{}
\begin{tabular}{|c|cc|c|cc|c|c|}
\hline
\multirow{2}{*}{\textbf{\begin{tabular}[c]{@{}c@{}}Alternatives (Density \\ per square mile)\end{tabular}}} & \multicolumn{3}{c|}{\textbf{\begin{tabular}[c]{@{}c@{}}GCM-t with the \\ estimable DOF\end{tabular}}} & \multicolumn{3}{c|}{\textbf{\begin{tabular}[c]{@{}c@{}}GCM-t with the \\ fixed DOF of 300\end{tabular}}} & \multirow{2}{*}{\textbf{\begin{tabular}[c]{@{}c@{}}t-stat for the difference \\ in mean elasticity\end{tabular}}} \\ \cline{2-7}
 & \textbf{Mean} & \textbf{Std. Dev.} & \textbf{t-stat} & \textbf{Mean} & \textbf{Std. Dev.} & \textbf{t-stat} &  \\ \hline
0-99 & 0.0076 & 0.0025 & 3.04 & 0.0129 & 0.0030 & 4.30 & 9.23 \\
100-499 & 0.0045 & 0.0025 & 1.80 & 0.0037 & 0.0024 & 1.54 & 4.08 \\
500 – 1499 & 0.0147 & 0.0016 & 9.19 & 0.0135 & 0.0030 & 4.50 & 27.24 \\
1500-2000 & -0.0037 & 0.0022 & -1.68 & -0.0135 & 0.0025 & -5.40 & 12.53 \\
2000+ & -0.0232 & 0.0026 & -8.92 & -0.0165 & 0.0015 & -11.00 & 60.78 \\ \hline
\end{tabular}
\end{adjustwidth}
\end{table}

\begin{table}[h!]
		\centering
		\caption{Sample of a choice situation in the discrete choice experiment \vspace{-.3cm}}
		\label{tab:6}
		\begin{adjustwidth}{1.5cm}{}
\begin{tabular}{|l|c|c|}
\hline
 & \textbf{Gasoline version} & \textbf{Electric version} \\ \hline
Purchase Price & \$19,000 & \$34,000 \\
Driving cost per 50 miles & \$5.5 per 50 miles & \$2.5 per 50 miles \\
Electric driving range &  & 250 miles \\
EV parking: charging time for 50 miles &  & 90 minutes per 50 miles \\
EV parking: charging type &  & On-street \\
EV parking: time to find space &  & 5 minutes \\
EC parking: monthly price &  & \$50 per month \\ \hline
\end{tabular}
\vspace{.5cm}
\\Given the 2 options above. which car would you buy? 
\begin{itemize}
    \item Gasoline version
    \item Electric version
    \item Neither
\end{itemize}
\end{adjustwidth}
\end{table}

\begin{table}[h!]
		\centering
		\caption{Descriptive statistics of the sample \vspace{-.3cm}}
		\label{tab:7}
		\begin{adjustwidth}{2.5cm}{}
	\begin{tabular}{|l|c|}
\hline
\textbf{Variables} & \textbf{Frequency/Average} \\ \hline
Married indicator & 36.45\% \\
Indicator for having children & 49.55\% \\
Indicator for working full time & 59.47\% \\
Indicator for holding master's or above degree & 17.70\% \\
Number of adults in household & 2.88 \\
Number of driving-license-holders in household & 2.03 \\
Number of vehicles in household & 1.70 \\ 
Indicator for owning hybrid electric vehicle & 4.17\%\\
Male indicator & 29.51\% \\
Hispanic indicator & 9.53\% \\
Walk score & 77.33 \\
Population density & 21.53 \\
Household annual vehicle miles traveled & 14883.60 \\  
\textbf{Race} & \\
African-American indicator & 24.38\% \\
Asian indicator & 3.76\% \\
Caucasian indicator & 62.06\% \\
\textbf{Age category} & \\
Baby-boomer indicator & 16.41\% \\
GenX indicator & 24.58\% \\
Millenial indicator & 57.65\% \\\hline
\end{tabular}
\end{adjustwidth}
\end{table}

\begin{table}[h!]
		\centering
		\caption{Empirical study: comparison of parameter estimates of GCM-t and GCM-N (t-value in parenthesis) \vspace{-0.3cm}}
		\label{tab:8}
		\begin{adjustwidth}{-1.5cm}{}
		\resizebox{1.2\textwidth}{!}{
\begin{tabular}{|l|cccc|cccc|}
\hline
\textbf{Model fit Statistics} & \multicolumn{4}{c}{\textbf{GCM-t Model}} & \multicolumn{4}{|c|}{\textbf{GCM-N Model}} \\  \hline 
Sample size & \multicolumn{4}{c}{12336} & \multicolumn{4}{|c|}{12336} \\
Number of parameters & \multicolumn{4}{c}{39} & \multicolumn{4}{|c|}{42} \\
Loglikelihood & \multicolumn{4}{c}{-22725.81} & \multicolumn{4}{|c|}{-22929.19} \\
Bayesian Information Criterion & \multicolumn{4}{c}{45611.18} & \multicolumn{4}{|c|}{46030.21} \\
Trace of covariance matrix & \multicolumn{4}{c}{1.06} & \multicolumn{4}{|c|}{3.43} \\ \hline
\multirow{2}{*}{\textbf{Parameters}} & \textbf{Continuous} & \multicolumn{3}{c}{\textbf{Unordered}} & \textbf{Continuous} & \multicolumn{3}{c|}{\textbf{Unordered}} \\ \cline{2-9} 
 & \textbf{\begin{tabular}[c]{@{}c@{}}Household vehicle \\ miles traveled \end{tabular}} & \textbf{Gasoline} & \textbf{Electric} & \textbf{Opt-out} & \textbf{\begin{tabular}[c]{@{}c@{}}Household vehicle \\  miles traveled \end{tabular}} & \textbf{Gasoline} & \textbf{Electric} & \textbf{Opt-out} \\ \hline
Intercept & \begin{tabular}[c]{@{}c@{}}1.983\\ (81.22)\end{tabular} &  & \begin{tabular}[c]{@{}c@{}}-1.755\\ (-5.69)\end{tabular} & \begin{tabular}[c]{@{}c@{}}-4.229\\ (-9.82)\end{tabular} & \begin{tabular}[c]{@{}c@{}}1.856\\ (34.64)\end{tabular} &  & \begin{tabular}[c]{@{}c@{}}-1.599\\ (-5.66)\end{tabular} & \begin{tabular}[c]{@{}c@{}}-5.027\\ (-7.83)\end{tabular} \\ \hline
\multicolumn{9}{|l|}{\textbf{Demographic Variables}} \\ \hline
Married indicator & \begin{tabular}[c]{@{}c@{}}-0.059\\ (-4.38)\end{tabular} &  &  &  & \begin{tabular}[c]{@{}c@{}}-0.105\\ (-6.19)\end{tabular} &  & \begin{tabular}[c]{@{}c@{}}-0.033\\ (-1.10)\end{tabular} &  \\
Indicator for having children &  &  & \begin{tabular}[c]{@{}c@{}}-0.244\\ (-7.80)\end{tabular} &  & \begin{tabular}[c]{@{}c@{}}-0.030\\ (-4.50)\end{tabular} &  & \begin{tabular}[c]{@{}c@{}}-0.227\\ (-7.09)\end{tabular} &  \\
Indicator for working full time & \begin{tabular}[c]{@{}c@{}}0.174\\ (13.38)\end{tabular} &  &  &  & \begin{tabular}[c]{@{}c@{}}0.136\\ (9.16)\end{tabular} &  &  &  \\
\begin{tabular}[c]{@{}l@{}}Indicator for holding master's \\ or above degree \end{tabular}  &  &  & \begin{tabular}[c]{@{}c@{}}0.144\\ (3.94)\end{tabular} &  & \begin{tabular}[c]{@{}c@{}}0.025\\ (1.14)\end{tabular} &  & \begin{tabular}[c]{@{}c@{}}0.119\\ (3.44)\end{tabular} &  \\
Number of adults in household &  &  &  &  & \begin{tabular}[c]{@{}c@{}}0.109\\ (15.05)\end{tabular} &  & \begin{tabular}[c]{@{}c@{}}0.049\\ (4.15)\end{tabular} &  \\
\begin{tabular}[c]{@{}l@{}}Number of driving license holders \\ in household\end{tabular} & \begin{tabular}[c]{@{}c@{}}0.055\\ (6.86)\end{tabular}  &  & \begin{tabular}[c]{@{}c@{}}0.076\\ (3.75)\end{tabular} &  & \begin{tabular}[c]{@{}c@{}}-0.033\\ (-2.91)\end{tabular} &  &  &  \\
Number of vehicles in household &  &  & \begin{tabular}[c]{@{}c@{}}-0.076\\ (-3.78)\end{tabular} &  & \begin{tabular}[c]{@{}c@{}}0.013\\ (1.29)\end{tabular} &  & \begin{tabular}[c]{@{}c@{}}-0.068\\ (-4.14)\end{tabular} &  \\
\begin{tabular}[c]{@{}l@{}}Indicator for owning hybrid \\ electric vehicle\end{tabular}   &  &  & \begin{tabular}[c]{@{}c@{}}0.580\\ (8.12)\end{tabular} &  &  &  & \begin{tabular}[c]{@{}c@{}}0.522\\ (8.02)\end{tabular} &  \\
Male indicator &  &  &  &  &  &  & \begin{tabular}[c]{@{}c@{}}0.081\\ (2.85)\end{tabular} &  \\
Hispanic indicator &  &  & \begin{tabular}[c]{@{}c@{}}-0.139\\ (-2.80)\end{tabular} &  & \begin{tabular}[c]{@{}c@{}}-0.061\\ (-2.62)\end{tabular} &  & \begin{tabular}[c]{@{}c@{}}-0.119\\ (-2.77)\end{tabular} &  \\
\textbf{Race (base: Caucasian)} &  &  &  &  &  &  &  &  \\
African-American indicator &  &  &  &  & \begin{tabular}[c]{@{}c@{}}0.159\\ (9.62)\end{tabular} &  & \begin{tabular}[c]{@{}c@{}}-0.071\\ (-2.25)\end{tabular} &  \\
Asian indicator &  &  &  &  & \begin{tabular}[c]{@{}c@{}}-0.102\\ (-2.80)\end{tabular} &  & \begin{tabular}[c]{@{}c@{}}0.274\\ (4.31)\end{tabular} &  \\
\textbf{Age category (base: Millennial)} &  &  &  &  &  &  &  &  \\
Baby boomer indicator &  &  & \begin{tabular}[c]{@{}c@{}}-0.473\\ (-10.30)\end{tabular} &  &  &  & \begin{tabular}[c]{@{}c@{}}-0.436\\ (-10.10)\end{tabular} &  \\
GenX indicator &  &  & \begin{tabular}[c]{@{}c@{}}-0.207\\ (-5.51)\end{tabular} &  &  &  & \begin{tabular}[c]{@{}c@{}}-0.189\\ (-5.60)\end{tabular} &  \\ \hline
\multicolumn{9}{|l|}{\textbf{Environmental variables}} \\ \hline
Walk-score &  &  & \begin{tabular}[c]{@{}c@{}}0.635\\ (6.46)\end{tabular} &  & \begin{tabular}[c]{@{}c@{}}0.103\\ (1.36)\end{tabular} &  & \begin{tabular}[c]{@{}c@{}}0.538\\ (6.17)\end{tabular} &  \\
Population density of neighborhood & \begin{tabular}[c]{@{}c@{}}0.004\\ (5.60)\end{tabular}  &  &  &  & \begin{tabular}[c]{@{}c@{}}0.003\\ (2.93)\end{tabular} &  &  &  \\ \hline
\multicolumn{9}{|l|}{\textbf{Alternative specific variables}} \\ \hline
Price (in \$1,000) &  & \begin{tabular}[c]{@{}c@{}}-0.037\\ (-6.41)\end{tabular} & \begin{tabular}[c]{@{}c@{}}-0.019\\ (-3.48)\end{tabular} &  &  & \begin{tabular}[c]{@{}c@{}}-0.037\\ (-6.77)\end{tabular} & \begin{tabular}[c]{@{}c@{}}-0.021\\ (-4.13)\end{tabular} &  \\
Operating cost per 50 miles (in \$) &  & \begin{tabular}[c]{@{}c@{}}-0.284\\ (-7.16)\end{tabular} & \begin{tabular}[c]{@{}c@{}}-0.426\\ (-6.97)\end{tabular} &  &  & \begin{tabular}[c]{@{}c@{}}-0.245\\ (-6.57)\end{tabular} & \begin{tabular}[c]{@{}c@{}}-0.328\\ (-6.18)\end{tabular} &  \\
Log of driving range (in 100 miles)  &  &  & \begin{tabular}[c]{@{}c@{}}0.604\\ (8.29)\end{tabular} &  &  &  & \begin{tabular}[c]{@{}c@{}}0.481\\ (7.50)\end{tabular} &  \\
\begin{tabular}[c]{@{}l@{}}Electric vehicle charging time \\ (in hours) \end{tabular} &  &  & \begin{tabular}[c]{@{}c@{}}-0.215\\ (-7.37)\end{tabular} &  &  &  & \begin{tabular}[c]{@{}c@{}}-0.169\\ (-6.37)\end{tabular} &  \\
\begin{tabular}[c]{@{}l@{}}Electric vehicle parking search time \\ (in minutes)\end{tabular}   &  &  & \begin{tabular}[c]{@{}c@{}}-0.015\\ (-3.00)\end{tabular} &  &  &  & \begin{tabular}[c]{@{}c@{}}-0.013\\ (-2.99)\end{tabular} &  \\
\begin{tabular}[c]{@{}l@{}}Monthly electric vehicle parking \\ cost (in \$100)\end{tabular} &  &  &  \begin{tabular}[c]{@{}c@{}}-0.224\\ (-3.05)\end{tabular} &  &  &  & \begin{tabular}[c]{@{}c@{}}-0.282\\ (-4.39)\end{tabular} &  \\ \hline
\multicolumn{9}{|l|}{\textbf{Structural effect}} \\ \hline
\begin{tabular}[c]{@{}l@{}}Household vehicle miles traveled \\ (in 1000 miles)\end{tabular} &  &  & \begin{tabular}[c]{@{}c@{}}-0.019\\ (-6.77)\end{tabular} &  &  &  & \begin{tabular}[c]{@{}c@{}}-0.007\\ (-4.85)\end{tabular} &  \\ \hline
\end{tabular}}
\end{adjustwidth}
\end{table}

\begin{table}[h!]
		\centering
		\caption{Degree-of-freedom specification results in GCM-t model (t-value in parenthesis) \vspace{-.3cm}}
		\label{tab:9}
		\begin{adjustwidth}{3.5cm}{}
\begin{tabular}{|l|c|}
\hline
\textbf{Parameter} & \textbf{\begin{tabular}[c]{@{}c@{}}Estimates \\ (T-value)\end{tabular}} \\ \hline
Intercept & \begin{tabular}[c]{@{}c@{}}1.191\\ (4.67)\end{tabular} \\
Male indicator & \begin{tabular}[c]{@{}c@{}}0.202\\ (2.25)\end{tabular} \\
Married indicator & \begin{tabular}[c]{@{}c@{}}0.128\\ (1.46)\end{tabular} \\
Indicator for holding master's or above degree & \begin{tabular}[c]{@{}c@{}}0.182\\ (1.67)\end{tabular} \\
Number of adults in household & \begin{tabular}[c]{@{}c@{}}-0.165\\ (-5.32)\end{tabular} \\
Number of driving license holders in household & \begin{tabular}[c]{@{}c@{}}0.132\\ (2.65)\end{tabular} \\
Walk score & \begin{tabular}[c]{@{}c@{}}0.372\\ (1.30)\end{tabular} \\ 
\textbf{Race (base: Caucasian)} &   \\
African-American indicator & \begin{tabular}[c]{@{}c@{}}-0.396\\ (-4.64)\end{tabular} \\
Asian indicator & \begin{tabular}[c]{@{}c@{}}-0.309\\ (-1.48)\end{tabular} \\
\hline
\end{tabular}
\\ \footnotesize{\textbf{Note 1:} We also estimate GCM-t with constant DOF, and the estimated DOF is 3.37.\\
\textbf{Note 2:} Other results of the constant-DOF model are available upon request.} 
\end{adjustwidth}
\end{table}

\begin{table}[h!]
		\centering
		\caption{Change in Probability of alternatives due to 1\% reduction in parking-cost of electric vehicle \vspace{-.3cm}}
		\label{tab:10}
		\begin{adjustwidth}{-1cm}{}
\begin{tabular}{|cc|ccc|ccc|}
\hline
\multirow{2}{*}{\textbf{DOF lower limit}} & \multirow{2}{*}{\textbf{DOF upper limit}} & \multicolumn{3}{c|}{\textbf{GCM-t model}} & \multicolumn{3}{c|}{\textbf{GCM-N model}} \\  \cline{3-8} 
 &  & \textbf{Gasoline} & \textbf{Electric Vehicle} & \textbf{opt-out} & \textbf{Gasoline} & \textbf{Electric Vehicle} & \textbf{opt-out} \\ \hline
1.27 & 2.37 & -0.0002 & 0.0001 & 0.0000 & -0.0003 & 0.0003 & 0.0000 \\
2.37 & 2.83 & -0.0002 & 0.0002 & 0.0000 & -0.0003 & 0.0002 & 0.0000 \\
2.83 & 3.16 & -0.0002 & 0.0002 & 0.0000 & -0.0003 & 0.0003 & 0.0000 \\
3.16 & 3.53 & -0.0002 & 0.0002 & 0.0000 & -0.0003 & 0.0002 & 0.0000 \\
3.53 & 3.86 & -0.0002 & 0.0002 & 0.0000 & -0.0003 & 0.0003 & 0.0000 \\
3.86 & 4.16 & -0.0002 & 0.0002 & 0.0000 & -0.0003 & 0.0003 & 0.0000 \\
4.16 & 4.40 & -0.0002 & 0.0002 & 0.0000 & -0.0004 & 0.0003 & 0.0000 \\
4.40 & 4.83 & -0.0002 & 0.0002 & 0.0000 & -0.0004 & 0.0003 & 0.0000 \\
4.83 & 5.33 & -0.0003 & 0.0002 & 0.0000 & -0.0004 & 0.0003 & 0.0000 \\
5.33 & 9.32 & -0.0003 & 0.0002 & 0.0000 & -0.0004 & 0.0003 & 0.0000 \\ \hline
\end{tabular}
\end{adjustwidth}
\end{table}	

\begin{table}[h!]
		\centering
		\caption{Change in Probability of alternatives due to 25\% reduction in parking-cost of electric vehicle \vspace{-.3cm}}
		\label{tab:11}
		\begin{adjustwidth}{-1cm}{}
\begin{tabular}{|cc|ccc|ccc|}
\hline
\multirow{2}{*}{\textbf{DOF lower limit}} & \multirow{2}{*}{\textbf{DOF upper limit}} & \multicolumn{3}{c|}{\textbf{GCM-t model}} & \multicolumn{3}{c|}{\textbf{GCM-N model}} \\  \cline{3-8} 
 &  & \textbf{Gasoline} & \textbf{Electric Vehicle} & \textbf{opt-out} & \textbf{Gasoline} & \textbf{Electric Vehicle} & \textbf{opt-out} \\ \hline
1.27 & 2.37 & -0.0050 & 0.0038 & -0.0005 & -0.0087 & 0.0064 & -0.0004 \\
2.37 & 2.83 & -0.0054 & 0.0041 & -0.0006 & -0.0088 & 0.0061 & -0.0006 \\
2.83 & 3.16 & -0.0052 & 0.0044 & -0.0005 & -0.0084 & 0.0068 & -0.0005 \\
3.16 & 3.53 & -0.0053 & 0.0043 & -0.0004 & -0.0084 & 0.0062 & -0.0004 \\
3.53 & 3.86 & -0.0056 & 0.0050 & -0.0006 & -0.0089 & 0.0073 & -0.0005 \\
3.86 & 4.16 & -0.0057 & 0.0051 & -0.0005 & -0.0088 & 0.0070 & -0.0006 \\
4.16 & 4.40 & -0.0062 & 0.0053 & -0.0004 & -0.0094 & 0.0072 & -0.0006 \\
4.40 & 4.83 & -0.0060 & 0.0052 & -0.0007 & -0.0091 & 0.0070 & -0.0009 \\
4.83 & 5.33 & -0.0072 & 0.0053 & -0.0004 & -0.0104 & 0.0073 & -0.0005 \\
5.33 & 9.32 & -0.0069 & 0.0046 & -0.0005 & -0.0100 & 0.0064 & -0.0007 \\ \hline
\end{tabular}
\end{adjustwidth}
\end{table}

	\begin{table}[h!]
		\centering
		\caption{Ratio of GCM-t and GCM-N probabilities for chosen alternative for different DOF \vspace{-.3cm}}
		\label{tab:12}
		\begin{adjustwidth}{1cm}{}
\begin{tabular}{|cc|ccc|c|}
\hline
\textbf{DOF lower limit} & \textbf{DOF upper limit} & \textbf{Gasoline} & \textbf{Electric Vehicle} & \textbf{opt-out} & \textbf{Sample size} \\ \hline
1.27 & 2.37 & 0.96 & 0.96 & 1.91 & 1232 \\
2.37 & 2.83 & 0.97 & 1.00 & 1.33 & 1232 \\
2.83 & 3.16 & 0.99 & 0.96 & 1.30 & 1232 \\
3.16 & 3.53 & 0.98 & 1.00 & 1.18 & 1208 \\
3.53 & 3.86 & 1.01 & 0.97 & 1.44 & 1264 \\
3.86 & 4.16 & 1.00 & 1.00 & 0.98 & 1216 \\
4.16 & 4.40 & 1.01 & 1.00 & 0.77 & 1248 \\
4.40 & 4.83 & 1.02 & 1.00 & 0.87 & 1224 \\
4.83 & 5.33 & 1.02 & 1.01 & 0.96 & 1240 \\
5.33 & 9.32 & 1.03 & 1.00 & 0.80 & 1232 \\ \hline
\multicolumn{2}{|c|}{\textbf{Overall Average}} & \textbf{1.00} & \textbf{0.99} & \textbf{1.15} &  \\ \hline
\end{tabular}
\end{adjustwidth}
\end{table}

\begin{table}[h!]
		\centering
		\caption{Comparison of Brier scores in the five-fold cross-validation analysis \vspace{-.3cm}}
		\label{tab:rev2}
		\begin{adjustwidth}{.5cm}{}
\begin{tabular}{|c|c|cc|cc|cc|}
\hline
\textbf{} & \multirow{2}{*}{\textbf{Model}} & \multicolumn{2}{c|}{\textbf{Gasoline}} & \multicolumn{2}{c|}{\textbf{Electric}} & \multicolumn{2}{c|}{\textbf{Opt-out}} \\ \cline{3-8} 
\textbf{} &  & \textbf{Mean} & \textbf{Std. Dev.} & \textbf{Mean} & \textbf{Std. Dev.} & \textbf{Mean} & \textbf{Std. Dev.} \\ \hline
\multirow{2}{*}{In-Sample} & GCM-T & 5791.99 & 21.73 & 5480.91 & 22.03 & 2166.28 & 31.36 \\
 & GCM-N & 5803.92 & 15.83 & 5503.46 & 24.74 & 2194.23 & 33.31 \\ \hline
\multirow{2}{*}{Out-of-Sample} & GCM-T & 937.68 & 6.53 & 866.83 & 7.87 & 176.09 & 5.74 \\
 & GCM-N & 934.35 & 8.09 & 868.79 & 9.04 & 167.36 & 5.70 \\ \hline
\end{tabular}
\\ \footnotesize{\textbf{Note:} Brier Score $=\sum_{i} \sqrt{|r_i - p_i|}$ where $p_i$ is the choice probability obtained from the model and $r_i=1$ if the option was chosen by the decision-maker $i$, else it is 0.}
\end{adjustwidth}
\end{table}

\begin{table}[h!]
		\centering
		\caption{Covariance matrix (t-value in parenthesis) \vspace{-.3cm}}
		\label{tab:13}
		\begin{adjustwidth}{2.5cm}{}
\begin{tabular}{|ccc|c|ccc|}
\cline{1-3} \cline{5-7}
\textbf{GCM-t model} &  &  &  & \textbf{GCM-N model} &  &  \\ \cline{1-3} \cline{5-7} 
\begin{tabular}[c]{@{}c@{}}0.330\\ (40.42)\end{tabular} &  &  &  & \begin{tabular}[c]{@{}c@{}}0.638\\ (84.77)\end{tabular} &  &  \\
\begin{tabular}[c]{@{}c@{}}0.109\\ (7.12)\end{tabular} & \begin{tabular}[c]{@{}c@{}}1.0 \\ (fixed)\end{tabular} &  &  & \begin{tabular}[c]{@{}c@{}}0.109\\ (5.42)\end{tabular} & \begin{tabular}[c]{@{}c@{}}1.0 \\ (fixed)\end{tabular} &  \\
\begin{tabular}[c]{@{}c@{}}-0.020\\ (-1.31)\end{tabular} & \begin{tabular}[c]{@{}c@{}}0.250\\ (2.35)\end{tabular} & \begin{tabular}[c]{@{}c@{}}0.728\\ (3.16)\end{tabular} &  & \begin{tabular}[c]{@{}c@{}}0.029\\ (1.07)\end{tabular} & \begin{tabular}[c]{@{}c@{}}0.348\\ (1.58)\end{tabular} & \begin{tabular}[c]{@{}c@{}}2.790\\ (2.60)\end{tabular} \\ \cline{1-3} \cline{5-7} 
\end{tabular}
\end{adjustwidth}
\end{table}

\end{document}